\documentclass[a4paper,11pt,amsmath,amssymb]{book}

\usepackage[latin1]{inputenc}
\usepackage[T1]{fontenc}
\usepackage[english]{babel}
\usepackage{amsfonts}
\usepackage{amssymb}
\usepackage{amsmath}
\usepackage{amsthm}
\usepackage{graphicx}
\usepackage{hyperref} 
\def\be{\begin{equation}}
\def\ee{\end{equation}}
\def\bc{\begin{center}}
\def\ec{\end{center}}

\title{\textbf{A statistical mechanics approach\\ to autopoietic immune
networks}}
\author{Adriano Barra\footnote{Dipartimento di Fisica, Sapienza Universit\`a di Roma, Italy}  \ Elena Agliari\footnote{Dipartimento di Fisica, Universit\`a di Parma, Italy}}
\date{January $2010$}

\begin{document}
\maketitle

\tableofcontents

\chapter{Introduction}

\section{Preface}
The aim of this work is to try to bridge over theoretical immunology
and disordered statistical mechanics. Our long term hope is to
contribute to the development of a quantitative theoretical
immunology from which practical applications may stem.

In order to make theoretical immunology appealing to the
statistical physicist audience we are going to work out a research
article which, from one side, may hopefully act as a benchmark for
future improvements and developments,  from the other side, it is
written in a very pedagogical way both from a theoretical physics
viewpoint as well as from the theoretical immunology one.

Furthermore, we have chosen to test our model describing a wide
range of features of the adaptive immune response in only a paper:
this has been necessary in order to emphasize the benefit
available when using disordered statistical mechanics as a tool
for the investigation. However, as a consequence, each section is
not at all exhaustive and would deserve deep investigation: for
the sake of completeness, we restricted details in the analysis of
each feature with the aim of
introducing a self-consistent model.

\section{Immune system and statistical mechanics}

The purpose of the immune system is to detect and neutralize the
molecules, or cells, dangerous for the body (antigens, which could
be foreign invaders - e.g. viruses or bacteria - or deranged  - e.g.
cancerous - cells of the host), without damaging healthy cells
\cite{a11}. Despite of the evident differences, to accomplish its
function, the immune system exhibits properties analogous to the
nervous system \cite{a68}: it "learns" not to attack healthy cells
and it "develops a memory" of the pathogens encountered as time
goes by. In theoretical immunology there are two main strands to
explain the functioning of the immune system that ultimately
represent two approaches, \emph{reductionist} and \emph{systemic} for the
modeling of nature in general. In the first and most popular
approach, lymphocytes basically operate independently or, better,
the researcher focuses on the action of the single lymphocyte and
on the details of its interactions (i.e. internal cascade signals,
etc.) rather than on the global behavior of all the lymphocytes
interacting with each other. In the second approach, pioneered in
immunology by Elrich \cite{a8} and Jerne \cite{a44}, the immune
system is thought of as a whole and designed as a network of cells
stimulated to proliferate by the affinity interactions of their
exchanging antibodies (a functional idiotypic network \cite{a30}).
Interestingly, the two approaches are not incompatible but
complementary. While the former deals primarily with the response
to a stimulus, the latter allows to explain the ability to learn
and memorize of the immune system and the tolerance to low doses
of antigen \cite{a101}. In the past, the immune network theory has been investigated, although not
exhaustively, with disordered statistical mechanics
tools \cite{a35}. However, recent and deep advances in the field of
statistical mechanics of highly diluted networks
\cite{a22,a54,a81,a90,a93,a98} now allow to combine the two
viewpoints described above and to develop a unified and quantitative
theory. Indeed, reductionist and systemic approaches can be
recovered as special cases of null and not-negligible connectivity
respectively. This unification should be extremely promising from
a biological as well as mathematical point of view.

To sketch our viewpoint on the systemic approach we make the following parallel: the immunologist
wonders about the details, even ``hyperfine'', of the structure and
about the interaction of the lymphocyte with its ``particular
environment''. Analogously, the condensed matter physicist studies
the molecule of water in every detail, from the angle between
hydrogen atoms to the constructive interference between the
orbital of hydrogen and of oxygen. This speculation is fundamental, yet not exhaustive. In fact, from these details, which allow an
accurate description of the molecule, we are not able to deduce,
the "emerging properties of the network" of molecules (e.g. a
water-ice phase transition) as the control parameters (pressure
and temperature) are tuned. Indeed, these phenomena do not depend
significantly on the details of the molecule structure but rather
on the collective effects due to the interactions of large numbers
of these molecules and the study of such effects is just the goal
of statistical mechanics. Hence, within this framework, we want to
read the theory of the immune networks.

In a nutshell, statistical mechanics (for discrete
systems as the one considered here, i.e. Curie-Weiss theories \cite{a18,a23,a89}
and their complex generalizations \cite{a15,a19,a20,a97}) is a
powerful approach to this problem. Within the one-body interaction
with the external stimuli, it can describe the behavior of the single
clone (made up of a set of identical lymphocytes) with its coupled
antigen, while, with the two-body interaction, it can describe the
clone networks; in this way, interpolating between the two, we can
recover the two prototypes of theoretical immunology; furthermore
this approach is strongly based on probability theory and on physical
variational principles which allow to make the theory even
quantitative and predictive.

Turning to the object of application, immunology became so far one
of the most investigated field of science and the plethora of its
variegate scientific outcomes increases enormously year by year,
such that trials for a general unifying theory should be
attempted.

We start by introducing the \emph{one-body} theory and noticing that in immunology it corresponds to what we call a ``Burnet-like behavior'', then we extend our model including the
\emph{two-body} theory and show that it recovers what we call a ``Jerne-like
behavior''; as a natural consequence, we will show how this
extends the approach of Counthino-Varela for the systemic
self/non-self distinction \cite{a38,a39}. After these results, we
show how hysteresis, with its remanent magnetization, can play the
role of the generator of memory cells from plasma cells,
according to the Clonal Selection Theory \cite{a6,a7}.
Finally, we show how low- and high-dose tolerances, as well as the bell-shaped response, appear as emergent
features in our model, while in theoretical immunology analysis they are often postulated a-priori.

Even though not exhaustive, our model may act as an alternative
starting backbone for this field of research.

\section{Theoretical immunology through complex system glasses}

The immune system exhibits an extremely broad ensemble of characters, however we are
going to focus just on a subset of the whole system,
namely the one constituted by B lymphocytes and their related
immunoglobulins. Despite being a small part of the immune
system, it is the main constituent of the adaptive response and, basically, the core of the system; furthermore, similar
considerations may hold, with little modifications, even for the
T-killer response, ultimately the adaptive system as a whole (apart from the fundamental T-helper regulations, which from disordered
statistical mechanics viewpoint are closer to pure glassy models
and quite different from the scenario we propose here, see \cite{a35}).

In the following, no mention to any other element (neutrophils,
macrophages, APC, etc) will be made and we refer to specialized
textbooks for their introduction \cite{a11}, although their
knowledge is not a prerequisite for reading the manuscript.

Let us just sketch, for the sake of clearness, that we are
trying to model the B-core of the immune system (as well as the
T-killer core) as imitative, eliciting, models: Stimulation is
expressed by a firing lymphocyte towards its nearest neighbors
while suppression is expressed by a quiescent one. Hence, in our scheme,
it is not the sign of the coupling to establish the kind of interaction, i.e. positive for imitative interaction and negative for anti-imitative interaction, which here is always positive or zero, but the state of
the single lymphocyte itself.

Frustration surely is expected to appear in the immune system,
but, in our approach, it is encoded into the T-helper regulation of
the two core (B, T-killers) responses, which deserve another
work for its investigation.

\subsection{Clonal expansion and one body perspective}

The main constituents of an (adaptive) immune system are
B-lymphocytes (B-cells), together with the T-lymphocytes, and free
antibodies produced by B-cells. B-cells and T-cells have specific
protein molecules on their surfaces, called receptors. The
receptors of B-cells are antibodies (Immunoglobulin,
Ig), which can recognize and connect to antigens in order to
neutralize them. Finally, the purpose of killer T-cells is to
attack and kill infected or deranged cells. The receptors of B-
and T-cells have specific $3$-dimensional structures, called
``idiotypes''. A family of B-cells generated by a proliferating
B-cell are called ``clones''; a clone and the antibodies which it
produces have the same idiotypes.

In a healthy human body at rest, it is estimate that the total
number of "sentinel" clones generated from a single B-cell (the
amount of identical lymphocytes) is about $10^2$ to $10^4$, the
total number of clones amounts to some $10^{12}-10^{14}$ (such
that diverse clones are around $10^{10}-10^{12}$, and the number of
antibodies is about $10^{18}$; remarkably the amount of
epitopes/idiotopes belonging to a given antibody are present in a
smaller number, i.e. order of $10^2$.

When antigens enter the body, those clones which recognize it will
bind to it. Aided by helper T-cells, B-cells of an activated clone
will proliferate, becoming antibody producing cells. The latter
will secrete large numbers of free antibodies, which attach to
antigen, neutralize it, and trigger killer cells into action. The
above is (a part of the) "clonal selection theory" and has been
confirmed experimentally.

We notice that this approach, pioneered by Burnet \cite{a6}, takes
into account an enormous amount of different data, and absolutely
does not rely on interactions among different lymphocytes, as it
deals with the external antigen interaction with the immune system,
where the network works at a completely hidden level.

Edelman first realized how the principal features of Burnet
theory was its bridge over a description in terms of antibodies
and one in terms of cells, implicitly defining the first postulate
of immunology (which has been always verified so far) \cite{a40}:

Each clone of B-cells produces always the same antibody
(hyper-mutations apart which will not be discussed here - see for
instance \cite{a25,a26} - and should pave the way to learning
\cite{a24}), so a given clone $i$ may be composed by $M$ identical
lymphocytes each producing the same identical antibody and the
immune system is built by $N$ of these clones.

\subsection{Immunoglobulin network and two bodies perspective}

The idea of an internal network appeared early in immunology
\cite{a8}, and its concretization happened when Jerne, in the
$70$'s, suggested that each antibody must have several idiotopes
which are detected by other antibodies. Via this mechanism, an
effective network of interacting antibodies is formed, in which
antibodies not only detect antigens, but also function as
individual internal images of certain antigens and are themselves
being detected and acted upon. These network interactions provide
a "dynamical memory" of the immune system, by keeping the
concentrations of antibodies (especially those representing
encountered antigens) at appropriate levels. This can be
understood as follows: At a given time a virus is introduced in
the body and starts replication. As a consequence, at high enough
concentration, it is found by the proper B-lymphocyte counterpart:
let us consider, for simplicity, a virus as a string of
information (i.e. $1001001$). The complementary B-cell producing
the antibody Ig1, which can be thought of as the string
$0110110$ (the dichotomy of a binary alphabet in strings mirrors
the one of the electromagnetic field governing chemical bonds)
then will start a clonal expansion and will release high levels of
Ig1. As a consequence, after a while, another B-cell will meet
$0110110$ and, as this string never (macroscopically) existed
before, attacks it by releasing the complementary string $1001001$, that,
actually, is a "copy" (internal image) of the original virus but
with no DNA or RNA charge inside. The interplay among these keeps
memory of the past infection. However, in the $90$'s, the
network theory was considered to be strongly marginal: It did not
appear as a part of a whole although it gave an appealing
mechanism for the implementation of memory in the immune system.

\subsection{Tolerance, responsiveness and autoimmunity}

Beyond memory storage, another featured by the immune system is very
impressive: it is able to attack antigens but not host molecules
or cells. Immunologist name this ability as the distinction among
{\em self} and {\em non-self}: Self/Non-Self discrimination is of
fundamental importance as several disease may appear if it is
non-properly working (this is the case of auto-immune pathologies
\cite{a103}).

In a nutshell, following the classical vision and according to
(sometimes called reductionist \cite{a40}) antigen-driven view of
the immune system, newborn lymphocytes learn from the beginning
the difference among self and non-self (it is assumed the
existence of an a-priori learning in specific regions of the body
- i.e. thymus - where all the lymphocytes are made to interact with
self and all the responding ones are killed). As a consequence, the
presence of autoimmunity in the system is due to a non proper
elimination of those B-cells which, at their early stage, failed to learn
such a difference. This defines the allopoietic viewpoint.

It must be stressed that without a two-body
interaction, which makes possible the existence of a network, this
property can not be spread on the whole system and, indeed, we
must assume that each lymphocyte stores the whole required
information by itself, namely the reductionist viewpoint.

However, within the idiotypic network theory started by Jerne, the
emergence of an interaction network allows the following
speculations on autopoiesis due to Varela, Counthino and coworkers
\cite{a38,a39}: The mutual interaction among lymphocytes rules out the need of an a-priori learning for
these cells, as tolerance to self may turn out to be an emerging
property of the immune network thought of as a whole. In
fact, it is the modulation and the mutual influence among interacting
immunoglobulins (and their corresponding clones indirectly) that
imposes quiescence or responsiveness of the clones as a
consequence of a given stimulus, which may be "self" or "non-self"
as well. In other words, antibodies are randomly produced and, as a consequence,
may react against anything (their idiotopes form somehow
a "base" in a proper space), however clones producing
self-reacting antibodies are always taken quiescent, in such a way that
they can produce only low - but not zero - concentrations of Igs
\cite{a103}, by the interaction with the network of all the others. Indeed, we stress that experimentally low dose of self-antibodies are
commonly found in healthy bodies \cite{a34,a36}.

We finish this section reporting, at a minimal descriptive level,
other universal, basic features displayed by the immune system
\cite{a11,a35}:
\begin{itemize}

\item Low Dose Tolerance: the immune system does not attack
proteins (of whatever kind, antigens or self-proteins) if their
concentration is below a threshold, whose value strongly depend on
the particular protein itself.

\item High Dose Tolerance: the immune system does not attack them
either if their concentration is too high.

\item Bell Shaped Response: the typical form of the immune
response (which may vary in amplitude and length) is the so called
"Bell-Shaped Response".

\item Development of memory cells: During the fight against the
infection the immune system stores information of the encountered
antigen developing memory cells, which, in a possibly secondary
re-infection, produce more-specific antibodies and
in greater quantity.

\item Multi-attachment ability: when an antigen is encountered and a
"segment" of this is presented, thought APC, to the adaptive
responders, not only one, but rather an ensemble of antibodies is usually produced to fight the
infection.

\item Self vs non-self discrimination: cells or proteins
belonging to the body are never macroscopically attacked by the
immune system.

\end{itemize}

\chapter{The minimal model}

Conceptually we reserve this chapter to a derivation of a
Hamiltonian for the lymphocytes, which will play the role of the
spins in standard statistical-mechanics models. In the next
chapter we will focus on the immunoglobulin, which will build the
interaction matrix for the lymphocyte clones and this will define
the model. All the rest of the paper will show its features.
At the end we stress that we work out the theory paying attention
only at its finite $N$ - and finite $M$ - behavior, so to compare
with data.

\section{Antibodies as vectors in the base of idiotopes}
We want to formalize this scenario within a statistical mechanics
context where interacting antibodies ultimately reflect the
interaction among lymphocytes due to the one-to-one postulate
previously introduced: from a "field theory language" \cite{a104},
the antibodies are the "fields" that the lymphocytes produce for
interacting: these can interact both among themselves and directly
with the lymphocytes. As the ratio among the amount of antibodies
versus lymphocytes is much greater than one we focus primarily
on the  antibody-antibody interaction, how to extend
this to the former case is straightforward as lymphocytes display antibodies on their
external surface.

In order to get a network of Igs links, we relax the earlier
simplifying assumption of  "a perfect mirror of a mirror" for the
interacting Igs. In fact, we are going to consider interactions among antibodies
formed by idiotopes such that the better the matches among
idiotopes, the stronger the stimulus received by the respective clones
via their immunoglobulins.

We consider the most generic antibody as a chain made of by the
possible expression of $L$ idiotopes. The assumption that each
antibody can be thought of as a string of the same length is based
on two observations: the molecular weight for each Igs is very
accurately close to $15\cdot10^4$ and each idiotope on average is
large as each other ("all the gamma-globulins have structural
characteristic surprisedly similar" \cite{a9}).
\newline
As a consequence the $L$ idiotopes may act as eigenvectors,
\begin{eqnarray}
\nonumber
 \xi_1 &=& (1,0,0,...,0) \\
 \nonumber
 \xi_2 &=& (0,1,0,...,0) \\
 \nonumber
 && ... \\
 \nonumber
 \xi_L &=& (0,0,0,...,1). \\
\end{eqnarray}
They form an orthogonal base in the $L$-dimensional space of the
antibodies $\Upsilon$. A generic antibody $\xi^i$ can then be
decomposed as a linear combination of these eigenvectors
$\{\xi^{i} \} = \lambda_1^{i} \xi_1, \lambda_2^{i} \xi_2, ...,
\lambda_L^{i} \xi_L$, with $\lambda_{\mu}^i \in (0,1)$ accounting
for the expression $(1)$ of a particular $\mu^{th}$ idiotope on
the $i^{th}$ antibody or its lacking $(0)$.
\newline
In this way the earlier distinction among epitope and paratope
suggested by Jerne is avoided (as in several recent approaches to
theoretical immunology \cite{a4,a5,a10,a67}) and is translated
into a complementary  product that we will define sharply in the
next chapter.
\newline
Roughly speaking, within the previous example (see sec.$1.2.2$),
both the strings $(1001000),(1001001)$ are reactive with
$(0110110)$, but the second is better as it matches all the
entries. As a counterpart the strings with several differences in
idiotope/epitope linking (i.e. 0111110 in the same example) do not
match and the corresponding lymphocytes are disconnected in the
network they belong to (it is straightforward to understand that
there are no links inside the lymphocytes of the same clone,
namely they act paramagnetically among each other). The fact that
the interaction of two Ig's is stronger when their relative
strings are more complementary responds to the kind of interaction
among their proteic structures: protein-protein interactions are
dominated by weak, short-range non covalent forces which arise
when the geometry of the two proteins are complementary, whatever
structures are assumed.

This naturally enlarges the idea of "mirror of mirror" into an
affinity matrix $J_{ij} \geq 0$, which, although
described throughout in the next chapter, will be used now as the starting point of
the following speculation.

\section{One-body and two-body Hamiltonian}\label{sec:1B2B}

In this section we are going to introduce the "Hamiltonian" of
our system. The Hamiltonian $H$ encodes the interactions among
lymphocytes as well as the interactions among lymphocytes and the external antigens, providing a
measure for the ``energy" of the system. For the reader with no physics background we will summarize the
 key concepts of statistical mechanics and thermodynamics, directly
 applied in immunology, in the section (\ref{termodinamica}).

First of all, let us formalize the interactions taking place within the system.
We consider an ensemble of $M$ identical lymphocytes
$\sigma_i^{\alpha}$, $\alpha=1,...,M$, all belonging to the
$i^{th}$ clone and $N$ all different clones $i=1,...,N$.
\newline
In principle $M$, the size of available "soldiers" within a given
clone (in an healthy human body at rest), can depend by the clone
itself, such that $M \to M_i$. However, for the sake of
simplicity, we are going to use the same $M$ for all the clones,
at least in equilibrium and in the linear response regime.

If the match among antibodies had to be perfect for recognizing
each other, then in order to reproduce all possible antibodies
obtained by the $L$ epitopes, the immune system would need $N \sim
\mathcal{O}(2^L)$
 lymphocytes. Conversely, if we relax the hypothesis of the perfect
match, only a fraction of such quantity is retained to manage the
repertoire, such that we can define the following scaling among
lymphocytes and antibodies: \be\label{NL} N = f(L) \exp (\gamma
L), \ee where $\gamma \in [0,1]$ encodes for the ratio of the
involved lymphocytes (the order of magnitude) and $f(L)$ is a
generic rational monomial in $L$ for the fine tuning (as often
introduced in complex systems \cite{a102}, we will see that $f(L)
\sim \sqrt{L}$).

Interestingly, a far-from-complete system is consistent with the
fact that binding between antigens and antibodies can occur even
when the match is not perfect: experimental measurements showed
that the affinity among antibody and anti-antibody is of the order
 of the $65/70$ percent or more (but not $100\%$) \cite{a4,a6,a70,a75}. Furthermore the
experimental existence of more than one antibody responding to a
given stimulus (multiple attachment \cite{a3}) confirms the
statement.

We can think of each lymphocyte as a binary variable
$\sigma_i^{\alpha} \in \pm 1$ (where $i$ stands for the $i$th
clone in some ordering and $\alpha$ for the generic element in the
$i$ subset) such that when it assumes the value $-1$, it is
quiescent (low level of antibodies secretion) and when it is $+1$
it is firing (high level of antibodies secretion).

The ability of newborn lymphocytes to spontaneously secreting low
dose of its antibody (corresponding to its genotype) even when not
stimulated is fundamental in order to retain the network
equilibrium and can be deepen in \cite{a42}. We stress once again
that within our approach the upper bound of the available firing
lymphocytes is conserved $M \neq M(t)$ so the exponential growth
of a clone when expanding after the exposition to the external
antigen, is translated here in the growing response of a clone to
the external field by which, from a situation with almost all its
$M$ are in the state $\sigma_i=-1$ switches to a scenario with all $\sigma_i=+1$.

To check immune responses we need to introduce the $N$ order
parameters $m_i$ as local magnetizations \be m_i=
\frac{1}{M}\sum_{\alpha=1}^{M}\sigma_i^{\alpha}(t), \ee where $i$
labels the clone and $\alpha$ the lymphocyte inside the clone's
family; The vector of all the $m_i$'s is depicted as $\bold{m}$
and the global magnetization as the average of all the $m_i$ as
$\langle m \rangle = N^{-1}\sum_i^N m_i$.

It is important to stress that the magnetizations, which play the
role of the principal order parameters, account for the averaged
concentration of firing lymphocytes into the immune network, such
that as $m_i \in [-1,1]$ we can define the concentrations of the
firing $i^{th}$ lymphocytes as
\begin{equation} \label{eq:concetration}
c_i(t) \equiv \exp\left[ \tau
\frac{(m_i(t)+1 )}{2} \right], \ \ \tau = \log M.
\end{equation}
Note that the
concentration is not normalized and ranges over several orders of
magnitude, from $\mathcal{O}(10^{0})$ when no firing lymphocyte is present
up to $\mathcal{O}(10^{12}) \sim M$  when all the lymphocytes of the
$i$th clone are  firing. Strictly speaking, the quiescence of a
given clone is a collective state where $\sim 10^2/10^3$ clones
are present; this can be understood, within a thermodynamical
framework, relaxing the idea that the system works at
"zero-temperature" (that is not really physical), in fact, a small
amount of noise would change the quiescent concentration from
strictly $1$ to a slightly higher value.

Now, let us turn to the external field and start with the ideal
case of perfect coupling among a given antigen and its lymphocyte
counterpart: let us label $\bold{h^i}$ the antigen displaying a
sharp match with the $i$-th antibody, hence described by the
string $\xi_{h^i} = \bar{\xi}_i$. In general, for unitary concentration of the antigen, the coupling with an
arbitrary antibody $k$ is $h_k^i$.

Following classical statistical mechanics \cite{a18,a89}, the
interaction among the two can be described as \be H_1= -\sum_k^N
h^i_k m_k, \ee such that if we suppose that at the time $t$ the
only applied stimulus is the antigen $\bold{h^1}$, all clones
but $1$, namely $i=2,...,N$, remain quiescent: the interaction
term among the system and the stimulus is simply $H_1= - h^1_1 m_1$.
Note that within this Hamiltonian alone the immune system is at
rest apart from the clone $i=1$ which is responding to the external
offense and that if we apply contemporary two external antigens
$h_1(t), h_2(t)$, the response is the sum of the two responses.

Of course also the generic external input $\tilde{\bold{h}}$, stemming from the superposition of $L$ arbitrary elementary stimuli, can
be looked as the effect of a string $\tilde{\xi}$ which can written in the idiotype basis such that $
\tilde{\xi} = \sum_{i=1}^L \lambda_i  \; \xi_i$. Moreover, in order to account for the temporal dependence of the antigen concentration we introduce the variable $c(t)$
accounting for its load at the time $t$, such that, generically,
several lymphocytes attack it (we will quantify the response in
the next chapter), as commonly seen in the experiments \cite{a3}.

As we discussed, it is reasonable to believe that all the
immunoglobulins have the same length $L$, on the other hand this is not
obvious for antigens which may arrive from different organisms and
places, such that their interactions with the immune system may be
different. In a nutshell, referring the reader to specific
textbook, let us only remark that Antigen
Presenting Cells, which are immune agents with the role of
presenting the antigens to the lymphocytes, before accounting for
these meetings, desegregates the enemies in pieces of "information
length" of order $L$ and put them on the proper surface
\cite{a11}.

So far we introduced the (reductionist) one-body theory, whose
"Hamiltonian"  is encoded into the expression $H_1$.
If we now take into account a "network" of clones we should
include their interaction term $H_2$. Coherently with $H_1$ we can
think at
\begin{equation} \label{eq:H_2}
H_2= -N^{-1} \sum_{i<j}^{N,N} J_{ij} m_i m_j.
\end{equation}

As anticipated, the Hamiltonian is the average of the "energy"
inside the system and thermodynamic prescription is that system
tries to minimize it. As a consequence, assuming $J_{ij}\geq 0$,
the energies are lower when their constituents behave in the same
way.
For $H_2$, two generic clones $i$ and $j$ in mutual interactions, namely $J_{ij}
> 0$, tend to imitate one another (i.e. if $i$ is quiescent, it
tries to make $j$ quiescent as well -suppression-, while if the
former is firing it tries to make firing even the latter
-stimulation-, and symmetrically $j$ acts on $i$).

It is natural to assume $J_{ij}$ as the affinity matrix: it
encodes how the generic $i$ and $j$ elements are coupled together
such that its high positive value stands for an high affinity
among the two. The opposite being the zero value, accounting for
the missing interaction.

If we consider the more general Hamiltonian $H=H_1+H_2$ we
immediately see that in the case of $J_{ij}=0$ for all $i,j$ we
recover the pure one-body description and the antigen-driven
viewpoint alone.  Different ratio among
 the weighted connectivity $w_i = \sum_j J_{ij}$ and $h_i$ will interpolate, time by time,
among two limits for each clone $i$.

\section{Approaching a statistical mechanics formulation}

In this section we introduce the basic principles of stochastic
dynamics for the evolution to equilibrium statistical mechanics of
the system we are interested in. Even though for discrete systems
two kinds of dynamics are available, parallel and sequential, we
are going to deepen only the latter as it is the one we will
implement in this work.

The argument is well known and several mathematical textbooks are
available \cite{a24,a33} (the expert reader may safely skip to
Sec. ($2.3.3$)), however, for the sake of completeness we briefly
summarize the fundamental steps directly implementing them into
the immunological framework we model.

\subsection{Master equations and Markov process}

We saw that the average behavior of the generic $i$th clone is
expressed via $m_i=M^{-1}\sum_{\alpha=1}^M \sigma_i^{\alpha}$.
The interactions among the clones and the external stimuli are
encoded into the Hamiltonian as follows \be\label{model}
H(\sigma;J)= N^{-1}\sum_{i<j}^{N,N} J_{ij}m_i m_j + \sum_i^N \tilde{h}_i
m_i. \ee Note that there are no interactions among lymphocytes
belonging to the same clone $(J_{ii} \equiv 0)$.

For the moment there is no need to define explicitly the topology of the underlying immune network  as the scheme applies in full
generality. We only stress that $J_{ij}$ is quenched, i.e. it does
not evolve with time, or at least it evolves on slower timescales
w.r.t. the ones involved by the $\sigma$'s (this ultimately
reflects the difference among which genotype and phenotype evolve
\cite{a105} and is the usual approach in closer context, i.e.
neural networks \cite{a14}). $J_{ij}$ plays the role of the
affinity matrix and can be thought of as a weighted symmetric
adjacency matrix \cite{a17}.
\newline
Once defined the field $\tilde{h}_i$ acting on the generic $i$th
clone at time $t$, the state of the system at this time is given
as the average of all its building lymphocytes, each of which
evolving time-step by time-step via a suitable dynamics.
\newline
Following standard disordered statistical mechanics approach
\cite{a94} we introduce the latter accordingly to
\begin{equation}\label{markov}
\sigma_i^{\alpha}(t+1) = sign \left( \tanh(\beta \varphi_i(t)) + \eta_i^{\alpha}(t)\right ),
\end{equation}
where $\varphi_i(t)$ is the overall stimulus felt by the $i$-th lymphocyte, given by
\begin{equation}
\varphi_i(t) = N^{-1} \sum_j^N J_{ij} m_j(t) + \tilde{h}_i(t),
\end{equation}
end the randomness is in the noise implemented via the random
numbers $\eta_i^{\alpha}$, uniformly drawn over the set $[-1,+1]$.
 $\beta$ rules the impact of this noise on the state
$\sigma_i^{\alpha}(t+1)$, such that for $\beta=\infty$ the process
is completely deterministic while for $\beta=0$ completely random
\footnote{The reader not acquainted with statistical mechanics may find
the above equations somewhat obscure, in which case, he may deepen
the link among the (rather unfamiliar) hyperbolic tangent and the
(much more familiar) logistic function usually introduced in
experimental data analysis in medicine \cite{a106} to realize the
freedom allowed beyond this choice.}.
\newline
In this framework the noise can be though of as made by several
different agents as the concentration of free radicals (which bind
randomly, decreasing the strength of the interactions) or the
concentration of fat molecules as cholesterol, which speeds down
the drift velocity for the lymphocytes decreasing the effective
connection among them (as a big difference with neural networks,
whose graphs have neurons as nodes and synapses as links, in
immune networks the graph underlying the model is intrinsically
dynamical as depends by the blood flow instead of static neuronal
tissues \cite{a51}).
\newline
In the sequential dynamics, under the assumption $M<<N$, at each
time step $t$ a single lymphocyte $l_t$ -randomly chosen among the
$M \times N$- is updated, such that its evolution becomes \be
P[\sigma^{\alpha}_{l_t}(t+1)]=\frac12 \big( 1 +
\sigma^{\alpha}_{l_t}(t)\tanh(\beta k_{l_t}(t))\big), \ee whose
deterministic zero-noise limit is immediately recoverable by
sending $\beta \to \infty$.
\newline
If we now look at the probability of the state at a given time
$t+1$, $P_{t+1}(\sigma)$, we get
\begin{eqnarray}\label{fokkerplanck} P_{t+1}(\sigma) &=& \frac1N \frac1M
\sum_{i,\alpha}^{N,M} \frac{1}{2}(1+ \sigma_i^{\alpha} \tanh(\beta
k_i (\sigma)))P_t(\sigma) \\ &+&  \frac1N \frac1M
\sum_{i,\alpha}^{N,M} \frac{1}{2}(1+ \sigma_i^{\alpha} \tanh(\beta
k_i (F_i^{\alpha}\sigma)))P_t(F_i^{\alpha}\sigma), \nonumber
\end{eqnarray} where we introduced the $M \times N$ flip-operators
$F_i^{\alpha}$, $i \in (1,...,N), \alpha \in (1,...,M)$, acting on
a generic observable $\phi(\sigma)$, as
\begin{eqnarray}
F_i^{\alpha}\Phi(\sigma_1^{\alpha},...,&+&\sigma_i^{\alpha}, ...,
\sigma_N^{\alpha},\sigma_1^{\beta},...,\sigma_N^{\beta},...,\sigma_1^{M},...,\sigma_N^{M})
=\\ =\Phi(\sigma_1^{\alpha},..., &-&\sigma_i^{\alpha}, ...,
\sigma_N^{\alpha},\sigma_1^{\beta},...,\sigma_N^{\beta},...,\sigma_1^{M},...,\sigma_N^{M}),\end{eqnarray}
such that we can write the evolution of the immune network as a
Markov process
\begin{eqnarray}
p_{t+1}(m) &=& \sum_{m'}W[m;m']p_t(m'),
\\ W[m;m'] &=& \delta_{m,m'} +
\frac1N \frac1M \sum_{i=1}^N \sum_{\alpha=1}^M \Big(
w_i^{\alpha}(F_i^{\alpha} m)\delta_{m, F m'} - w_i^{\alpha}(m)
\delta_{m,m'} \Big), \nonumber
\end{eqnarray}
with the transition rates $w_i^{\alpha}(m) = \frac12 [1-
\sigma_i^{\alpha} \tanh(\beta k_i)]$.

\subsection{Detailed balance and symmetric interactions}

If the affinity matrix is symmetric, detailed balance ensures that
there exists a stationary solution $P_{\infty}(m)$ such that
(restricting $\tilde{h}_i(t) \to \tilde{h}_i \in \mathbb{R} \
\forall i \in (1,...,N)$)
$$
W[m,m']P_{\infty}(m')= W[m',m]P_{\infty}(m).
$$
As the model fulfills this requisite, we want to deepen its
implication to detailed balance at least in the $\beta \to \infty$
limit, which makes the evolution a deterministic map holding for
each $\alpha \in (1,...,M)$.
\newline
This key feature ensures equilibrium and is worked out
specifically as
$$
\frac{e^{(\beta M^{-1} \sum_{\alpha=1}^M \sigma_i^{\alpha}
h_i(F_i^{\alpha} m))}P_{\infty}(F_i^{\alpha} m)}{\cosh(\beta
h_i(F_i m))} = \frac{e^{(-\beta M^{-1} \sum_{\alpha=1}^M
\sigma_i^{\alpha} h_i(m))}P_{\infty}(m)}{\cosh(\beta h_i(m))},
$$
which implies \be\label{MB} p_{\infty}(\sigma;J,h) \propto
\exp\Big(\frac{\beta}{2N}\sum_{ij}^N J_{ij}(\alpha)m_i m_j - \beta
\sum_i^N h_i m_i\Big) = \exp\Big(-\beta H_N(\sigma;J)\Big), \ee
namely the Maxwell-Boltzmann distribution \cite{a33} for the
Hamiltonian (\ref{model}).
\newline
In absence of external stimuli, and skipping here the
question about the needed timescales for "thermalization", the
system reaches an equilibrium that it is possible to work out
explicitly as we are going to show.
\newline
It is important to stress that the concept of equilibrium here has
nothing to share with a general equilibrium of the body. It simply
means equilibrium with respect to a particular choice of the
quenched antibody network (ultimately encoded into the $J_{ij}$).
\newline
For this detailed balanced system furthermore, the sequential
stochastic process (\ref{markov}) reduces to Glauber dynamics such
that the following simple expression for the transition rates
$W_i$ can be implemented \be W_i(m) = \Big( 1 + \exp(\beta \Delta
H(\sigma_i; J)) \Big)^{-1}, \ \  \Delta H(\sigma_i; J) = H(F_i m;
J) - H(m; J), \ee and will indeed be used in simulations through
the paper.

\subsection{Minimum energy and maximum entropy
principles}\label{termodinamica}

In the previous section we showed that, if the affinity matrix is
symmetric, so that detailed balance holds, the stochastic
evolution of our immune model approaches the Maxwell-Boltzmann
distribution (see eq.(\ref{MB})), which determines the
thermodynamic equilibria.

Thermodynamics describes the macroscopic features of the system
and statistical mechanics allows to obtain such a macroscopic
description starting by its microscopic foundation, so to say,
obtaining the global immune behavior by studying  the whole single
lymphocyte actions, and then, using Probability Theory (thanks to
the large numbers of  these agents), for averaging over the
ensemble with the weight encoded by $P_{\infty}(\sigma; J,h)$.
\newline
This scenario is achievable when both the "internal energy"
density of the system $u(\beta)$ and the "entropy" density
$s(\beta)$ are explicitly obtained:
\newline
In a nutshell, in physics, the energy of the system is defined as
the intensive average of the Hamiltonian $u(\beta) \sim
N^{-1}\langle H_N (\sigma;J,h) \rangle$, while the entropy
($s(\beta) = \sum_{\sigma} P_{\infty}(\sigma;J,h)\log
P_{\infty}(\sigma;J,h)$) is a measure of information  stored
inside the network:
\newline
When mapping from physics, we found implicitly paved  the bridge
with immunology; in fact, as suggested in \cite{a69}, the two
important "thermodynamic observables" of the immune system are its
{\em economy} and its {\em specificity}. Still following
\cite{a69} if  we assume that the immune system tries to maximize
its specificity (entropy in our parallel) and to minimize its cost
(energy in the same parallel) the way to statistical mechanics is
naturally merged.
\newline
Then the two prescription of minimizing the energy $u(\beta)$
(minimum energy principle) and maximizing the entropy $s(\beta)$
(second law of thermodynamics) with respect to the order
parameters give the full macroscopic behavior of the system.
\newline
We stress that the same approach, which may appear strange to
researchers not involved in complex statistical mechanics, holds
successfully in several different fields, from the closer neural
networks \cite{a14,a20}, to social or economic frameworks
\cite{a94,a95,a96} or computer science \cite{a91}.

To fulfil these prescriptions the free energy $f(\beta)= u(\beta)
- \beta^{-1}s(\beta)$ comes in help
 because, as it is straightforward to check, minimizing this
quantity corresponds to both maximizing entropy and minimizing
energy (at the given level of noise). Furthermore, and this is the
key bridge with stochastic processes, there is a deep relation
among statistical mechanics and their equilibrium measure
$P_{\infty}$, in fact \be P_{\infty}(\sigma; J,h) \propto
\exp(-\beta H(\sigma; J,h)), \ \ f(\beta) \equiv
\frac{-\mathbb{E}}{\beta N} \log \sum_{\sigma} \exp(-\beta
H_N(\sigma;J,h)). \nonumber \ee Hence, once the microscopic
interaction laws are encoded into the Hamiltonian, we can achieve
a specific expression for the free energy, from which we can
derive
\begin{eqnarray} u(\beta) &=& -\partial_{\beta} (\beta
f(\beta)) = N^{-1}\langle H(\sigma; J,h) \rangle, \\
s(\beta) &=& f(\beta) + \beta^{-1}\partial_{\beta}(\beta
f(\beta)).
\end{eqnarray}
The operator $\mathbb{E}$ that averages over the quenched
distribution of couplings makes the theory not "sample-dependent":
For sure each realization of the network will be different with
respect to some other in its details, but we expect that, after
sufficient long sampling, the averages and variances of observable
become unaffected by the details of the quenched variables.
\newline
The Boltzmann state is given by
\begin{equation}
\omega(\Phi(\sigma,J)) = \frac{1}{Z_{N}(\beta,a)}
\sum_{\{\sigma_N\}} \Phi(\sigma;J) e^{-\beta H_{N}(\sigma,J)},
\end{equation}
and the total average $\langle \Phi \rangle$ is defined as
\begin{equation}
\langle \Phi \rangle = \textbf{E}[\omega(\Phi(\sigma,J))].
\end{equation}
\medskip
\newline
It is easy to check that when the level of noise is too high
$(\beta \to 0)$, details of the Hamiltonian are unfelt by the
clones, so that each clone behave independently of each other.
\newline
At the contrary, when the level of noise is not too high and the
system may experience the rules encoded into the Hamiltonian, it
is easy to see that \be -\frac{\partial f_N(\beta,a)}{\partial
h_i} = \langle m_i \rangle \neq 0, \ \
-\frac{\partial{f_N(\beta,a)}}{\partial J_{ij}} = \langle m_i m_j
\rangle \neq 0. \ee Namely, the response of the system to the
$i^{th}$ external stimulus is encoded into the $i^{th}$ order
parameter (the concentration of the corresponding clone, i.e.
$1$-body term) and the response to the affinity matrix is encoded
into the correlation among different clones ($2$-bodies term).
\newline
Not surprisingly, the first description of the immune system,
early formalized by Burnet looking at the response to infections
deals with a one-body approach (it is the response to the external
field), while the description of the memory by Jerne deals with
the two body approach (able to store information into the network
by breaking the ergodicity \cite{a15}).

\chapter{Structure analysis} \label{ch:structure}
We consider a system made of by $N$ idiotypically different
clones, each denoted with an italic letter $i$ and associated to a
binary string $\xi_i$ of length $L$ encoding the specificity of
the antibody produced. Each entry $\mu$ of the $i$-th string is
extracted randomly according to the discrete uniform distribution
in such a way that $\xi_i^{\mu}=1$ ($\xi_i^{\mu}=0$) with
probability $1/2$; this choice corresponds to a minimal assumption
which can be possibly modified, yet preserving the structure of
our model, only quantitative results will change accordingly.

Now, given a couple of clones, say $i$ and $j$, the
$\mu$-th entries of the corresponding strings are said to be
complementary, iff $\xi_i^{\mu} \neq \xi_j^{\mu}$. Therefore, the
number of complementary entries $c_{ij} \in [0, L]$ can be written
as \be c_{ij} = \sum_{\mu = 1}^{L} [\xi_i^{\mu} (1 - \xi_j^{\mu})
+ \xi_j^{\mu} (1 - \xi_i^{\mu}) ] =  \sum_{\mu = 1}^{L}
[\xi_i^{\mu} +\xi_j^{\mu} -2\xi_i^{\mu} \xi_j^{\mu}]. \ee

The affinity between two antibodies and, more generally, among two
entities described by a vector in the idiotype basis, is expected
to depend on how much complementary their structures are. In fact,
the non-covalent forces acting among antibodies depend on the
geometry, on the charge distribution and on
hydrophilic-hydrophobic effects which give rise to an attractive
(repulsive) interaction for any complementary (non-complementary)
match. Consequently, in our model we assume that each
complementary / non-complementary entry yields an attractive /
repulsive contribute. In  general, attractive and repulsive
contributes can have different intensity and we quantify their
ratio with a parameter $\alpha \in \mathbb{R}^+$. Hence, we
introduce the functional $f_{\alpha,L}: \Upsilon \times \Upsilon
\rightarrow \mathbb{R}$ as \be \label{eq:affinity}
f_{\alpha,L}(\xi_i,\xi_j) \equiv
 [\alpha c_{ij} - (L-c_{ij})], \ee
 which provides a measure of how ``affine'' $\xi_i$ and $\xi_j$ are.
In principle, $f_{\alpha,L}(\xi_i,\xi_j)$ can range from $-L$
(when $\xi_i = \xi_j$) to $\alpha L$ (when all entries are
complemetary, i.e. $\xi_i = \bar{\xi}_j$). Now, when the repulsive
contribute prevails, that is $f_{\alpha,L} < 0$, the two antibodies do not see each other and
the coupling among the corresponding lymphocytes  $J_{ij} (\alpha,L)$ is set equal to
zero, conversely, we take $J_{ij} (\alpha,L)=
\exp[f_{\alpha,L}(\xi_i,\xi_j)] / \langle \tilde{J} \rangle_{\alpha,L}$, being
$\langle \tilde{J} \rangle_{\alpha,L}$ a proper normalizing factor (see Sec.~\ref{sec:dilution}).

Otherwise stated, nodes can interact pairwise according to a
coupling $J_{ij}(\alpha,L)$, which is defined as:
\begin{equation} \label{eq:J} J_{ij} (\alpha,L)
\equiv \Theta(f_{\alpha,L}(\xi_i,\xi_j))
\frac{\exp [f_{\alpha,L}(\xi_i,\xi_j)] }{ \langle \tilde{J} \rangle_{\alpha,L} },
\end{equation}
where $\Theta(x)$ is the
discrete Heaviside function returning $x$ if $x>0$, and $0$ if $x
\leq 0$.

Notice that the models introduced in \cite{bagley,farmer} also define the connections between antibodies, according to the number of matches between the chains representing antibodies.

Some remarks are in order here. The choice of an exponential law
connecting the affinity $f_{\alpha,L}(\xi_i,\xi_j)$ between two
strings and their relevant coupling $J_{ij}$ follows empirical
arguments, in fact, we expect the latter to depend sensitively on
how complementary the two strings are, possible spanning several
orders of magnitude. Notice that this choice is also consistent
with Parisi's intuition \cite{a35}. Moreover, the prefactor
$1/\langle \tilde{J} \rangle_{\alpha,L}$ is taken in such a way
that $J_{ij}$ \footnote{henceforth we will drop the dependence on
$\alpha$ and $L$, if not ambiguous} has finite (unitary) average
for any value of $\alpha$ and $L$. More precisely, $\langle
\tilde{J} \rangle_{\alpha,L}$ is just the average of
$\tilde{J}_{ij} (\alpha,L) \equiv
\Theta(f_{\alpha,L}(\xi_i,\xi_j)) \exp
[f_{\alpha,L}(\xi_i,\xi_j)]$, calculated over all possible
matchings between $\xi_i$ and $\xi_j$; this point will be deepened
in Sec.~\ref{sec:dilution}.

We conclude this section with a last remark. The idiotypic network describing
the mutual interaction among lymphocytes just stems from the
affinity between the relevant antibodies calculated according to
Eq. (\ref{eq:affinity}). In fact, as underlined before, (even when
quiescent) lymphocytes produce (low) quantities of specific
antibodies which constitute the means through which lymphocytes
interact with each others. To fix ideas, let us consider
lymphocytes denoted as $i$ and $j$, producing antibodies described
by $\xi_i$ and $\xi_j$. If the affinity between $i$ and $j$ is
positive, i.e. $J_{ij}>0$, when $i$ is firing it will secrete
large amount of specific antibodies eventually detected by $j$
which will be in turn stimulated to respond. Vice versa, if the
affinity is negative, i.e. $J_{ij}=0$, then $i$ and $j$ are not directly aware
of their reciprocal state as antibodies $\xi_i$ are
``transparent'' for $j$'s receptors. Therefore, the affinity
pattern between antibodies does generate the lymphocyte network. Indeed, as shown in the next section, Eq. (\ref{eq:affinity}) allows to completely describe the topology of the
emergent idyotipic network.

\begin{figure}
\begin{center}
\rotatebox{90}{\resizebox{3in}{!}{\includegraphics{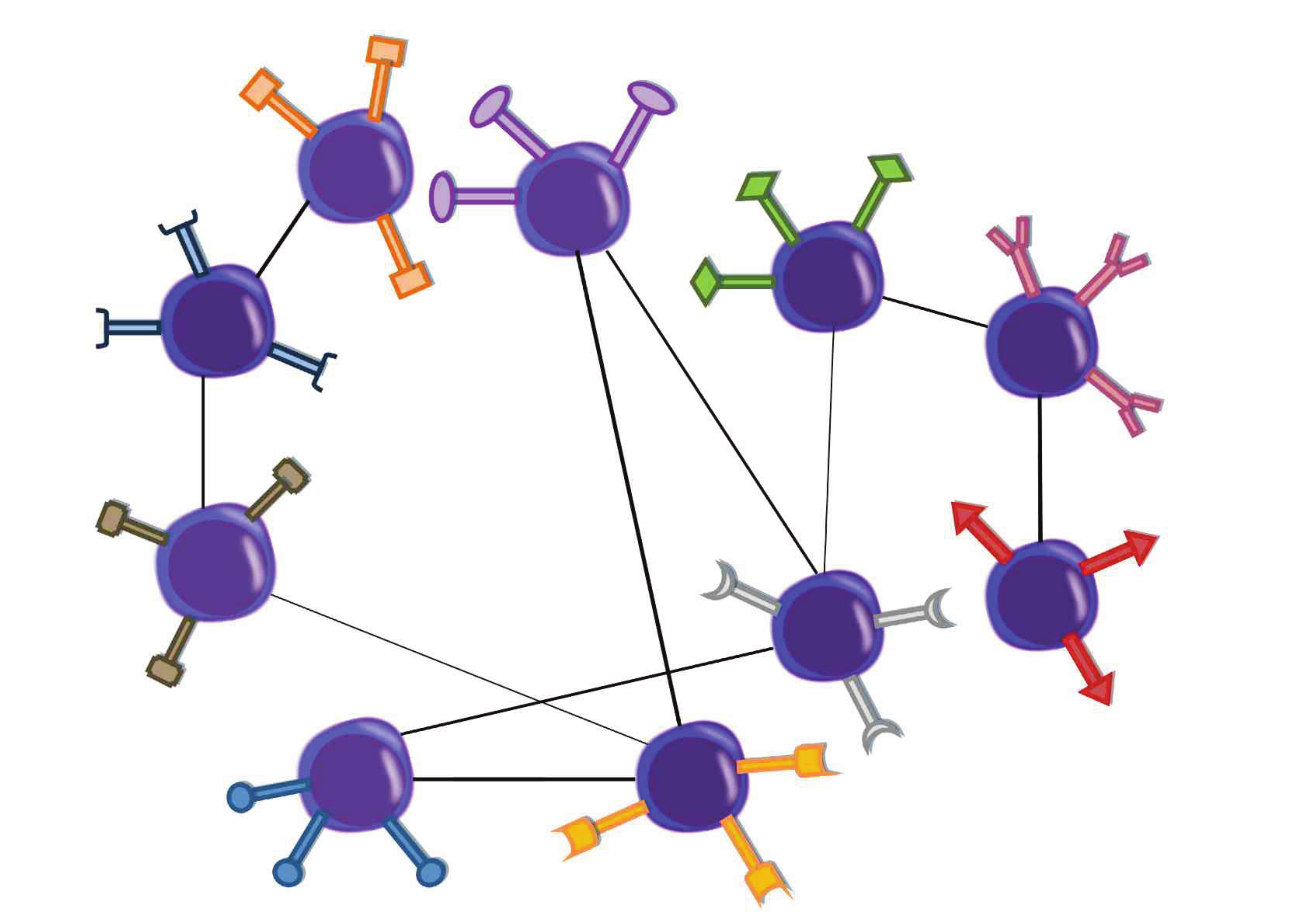}}}
\caption{Representation of the idyotipic network. Each clone is
represented by only one of its lymphocytes; the thickness of links denotes the strength of the corresponding coupling.}
\end{center}
\end{figure}

\section{Graphs} \label{sec:graphs}
The system of $N$ lymphocytes interacting pairwise with a coupling
$J_{ij}$ can be envisaged by means of a graph $\mathcal{G}$, whose
nodes represent lymphocytes and a link between them is drawn
whenever the pertaining coupling is positive. Before proceeding,
it is worth recalling that a generic graph $\mathcal{G}$ is
mathematically specified by the pair $\{V, \Gamma \}$ consisting
of a non-empty, countable set of points, $V$ joined pairwise by a
set of links $\Gamma$. The cardinality of $V$ is given by $|V| =
N$ representing the number of sites making up the graph, i.e. its
volume. From an algebraic point of view, a graph $\mathcal{G}=\{V,
\Gamma \}$ is completely described by its adjacency matrix
$\mathbf{A}$: Every entry of this off-diagonal, symmetric matrix,
corresponds to a pair of sites, and it equals one if and only if
this couple is joined by a link, otherwise it is zero. The number
of nearest-neighbors of the generic site $i$, referred to as
coordination number or degree, can be recovered as a sum of
adjacency matrix elements: $k_i = \sum_{j \in V} A_{ij}$.

In our model the graph describing the interaction among
lymphocytes is a random graph where links are drawn with
probability $p_{\alpha,L}$ which, in general, depends on the way
strings $\xi$'s are extracted and on the the way affinity
$f_{\alpha,L}$ is defined.

Here, due to the uniform distribution underlying the extraction of
$\xi$'s, we have that the probability that $\xi_i^{\mu}$ and
$\xi_j^{\mu}$ are complementary equals $1/2$ independently of $i,
j$ and $\mu$. Therefore, the probability that they display
$c_{ij}$ (hereafter simply $c$) complementary entries follows a
binomial distribution which reads off as \be \label{eq:P_c}
\mathcal{P}(c)= \left( \frac{1}{2} \right)^L \binom {L}{c}. \ee
Correspondingly, we have that lymphocytes $i$ and $j$ are
connected together, namely that $f_{\alpha,L}(\xi_i,\xi_j)>0$, when $c_{ij}(\alpha + 1) - L$ is positive (see Eq.~\ref{eq:affinity}) and this occurs with probability \be
\label{eq:p} p_{\alpha,L} = \sum_{c = \lfloor L/(\alpha+1) \rfloor
+1}^L \mathcal{P}(c), \ee where $\lfloor x \rfloor = \max \{ n \in
N | n \leq x \}.$

\begin{figure}[tb] \label{fig:P_k}
\begin{center}
\resizebox{0.7\columnwidth}{!}{\includegraphics{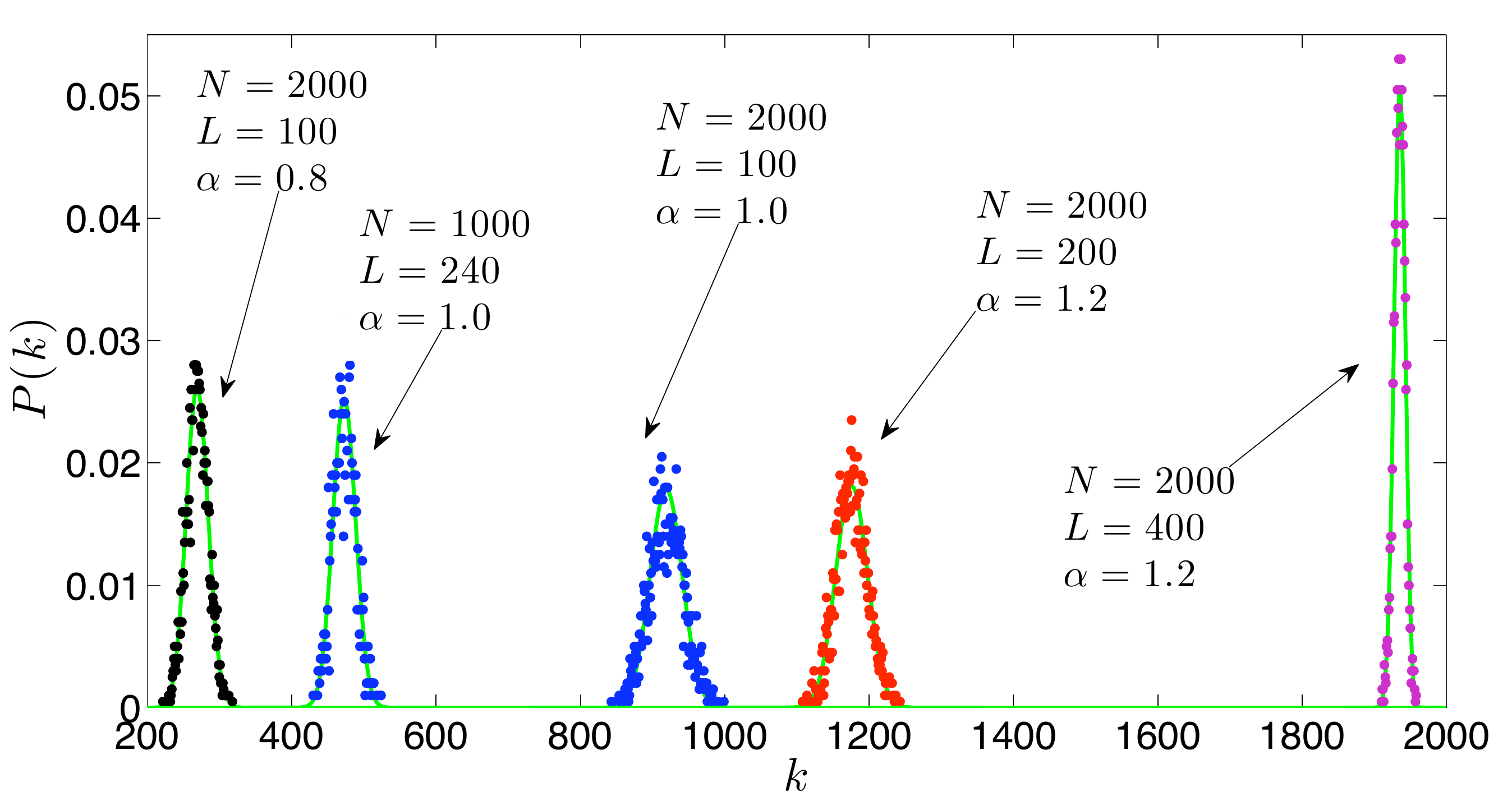}}
\caption{Degree distribution $P(k)$ for different values of $N$,
$L$ and $\alpha$; data from numerics ($\bullet$) and analytic
estimates (green lines), see Eq. (\ref{eq:P_k}), are compared.}
\end{center}
\end{figure}

The link probability $p_{\alpha,L}$ (see Eq. (\ref{eq:p})) is, at least for large $L$,
independent of the chosen couple, hence giving rise to an
Erd\"{o}s-Renyi graph $\mathcal{G}(N,p_{\alpha,L})$ \cite{a55}
characterized by a binomial degree distribution
\begin{equation}\label{eq:P_k}
P(k)=\binom{N}{k} p_{\alpha,L}^k (1-p_{\alpha,L})^k,
\end{equation}
representing the probability that a generic node has $k$ nearest
neighbors; the average degree follows as $\langle k \rangle
=p_{\alpha,L} (N-1)$, or, more simply, for $N$ large, we use
$\langle k \rangle = p_{\alpha,L} N$. In  Fig.~(\ref{fig:P_k})
we show the agreement between numerical data and analytic
estimates (Eq.~\ref{eq:P_k}) for the degree distribution $P(k)$;
different values of $N$, $L$, and $\alpha$ are considered.
The good agreement among data corroborates the analytical derivation based on the lack of correlation among links.

\section{Dilution}\label{sec:dilution}

The link probability $p_{\alpha,L}$ defined in Eq. (\ref{eq:p})
yields information about the dilution of the graph: the larger
$p_{\alpha,L}$, the more connected the graph. For instance, at given $L$, the
dilution of the graph can be controlled by properly tuning the
parameter $\alpha$.

As well known, by increasing the link probability from $0$
upwards, the (infinite) Erd\"{o}s-Renyi random graph undergoes a
percolation transition; namely there exists a critical link
probability $p_c$ such that when the link probability starts to
get larger than $p_c$ a so called ``giant component'', displaying
a size $\mathcal{O}(N)$, i.e. infinite in the thermodynamic limit, suddenly
appears \cite{a55}.

Indeed, the Erd\"{o}s-Renyi random graph $\mathcal{G}(N,p)$ can be
obtained from the complete graph of $N$ vertices, $K_N$, by
retaining each edge with probability $p$ and deleting it with
probability $1-p$, independently of all other edges. Analogously, the topology of the idiotypic network we are introducing in this work can be recovered from $K_N$ by conserving and erasing links with probability $p_{\alpha,L}$ and $1 - p_{\alpha,L}$, respectively.

The Molloy-Reed criterion for percolation \cite{a56} shows that,
for this model, a giant component exists if and only if $p$ is
larger than $1/N$. More precisely, setting $p=\langle k \rangle
/N$, if we denote with $C_1$ the largest component in
$\mathcal{G}(N,p)$ and with $|C_1|$ the cardinality of the set
$C_1$, then we have the following results (see \cite{a57}): if
$\langle k \rangle <1$, then with high probability (w.h.p.) $|C_1|
\sim \log N$, if $\langle k \rangle
>1$, then w.h.p. $|C_1| \sim N$, while if $\langle k \rangle = 1$,
the situation is more delicate and one can state $|C_1| \sim
N^{2/3}$.

Therefore, it is important to analyze in more details the behavior
of $p_{\alpha,L}$ as a function of $\alpha$ and $L$. For
$\alpha=1$ it is straightforward to see that $p_{1,L}=1/2$, due to
the symmetry of the distribution $\mathcal{P}(c)$ with respect to
$c=L/2$ \footnote{due to discreteness, for small $L$ this holds
rigorously only for $L$ odd, while for $L$ even $p_{1,L}$
approaches $1/2$ from below as $L$ gets larger.}.

More generally, for large $L$ we can adopt a continuous
description and write $p_{\alpha,L}$ (see Eq. (\ref{eq:p})) as
\begin{eqnarray} \label{eq:p_alfa_L}
p_{\alpha,L} &\approx& \int_{L/(\alpha +1)}^L \mathcal{P}(c) \, dc \approx   \int_{L/(\alpha +1)}^L \sqrt{\frac{2} {\pi L}} \,
e^{-\frac{(c-L/2)^2}{L/2}} \; dc\\  \label{eq:p_alfa_L2}
&=&  \frac{1}{2} \left[
\mathrm{Erf} \left( \sqrt{\frac{L}{2}} \right) - \mathrm{Erf}
\left( \frac{(1-\alpha)}{(1+\alpha) } \sqrt{\frac{L}{2}} \right)
\right],
\end{eqnarray}
where we replaced the distribution $\mathcal{P}(c)$ with the
normal distribution, having mean $L/2$ and variance $L/4$; in
fact, for $L$ large enough, the skew of the distribution
$\mathcal{P}(c)$ is not too great and we can approximate the
binomial distribution by the normal distribution \cite{a51}. From Eq.~\ref{eq:p_alfa_L} we can calculate the
derivative of $p_{\alpha,L}$ with respect to $\alpha$, which reads off as
\begin{equation}\label{eq:Dp_alfa_L}
\frac{\partial p_{\alpha,L} }{\partial \alpha} \approx
\sqrt{ \frac{2 L}{\pi} } \frac{1}{(1+\alpha)^2} e^{-\frac{L}{2} \tilde{\alpha}^2},
\end{equation}
where we called $\tilde{\alpha} \equiv (1 - \alpha) / (1 + \alpha)$.

We now turn to the coupling strength introduced in Eq.
(\ref{eq:J}) and we notice that we can write $\tilde{J}_{ij}=
\exp[ c_{ij}(\alpha+1)-L]$, whenever $c_{ij}>L/(\alpha+1)$,
otherwise $\tilde{J}_{ij}=0$. Hence, its mean value, averaged over
all possible matchings between two binary strings, can be written
as (see Eqs. (\ref{eq:p}),(\ref{eq:J})):
\begin{eqnarray}\label{eq:J_alfa_L}
\langle \tilde{J} \rangle_{\alpha,L} &\approx&  \int_{L/(\alpha+1)}^L
e^{c(\alpha+1)-L}\, \sqrt{\frac{2} {\pi L}} \,
e^{-\frac{(c-L/2)^2}{L/2}}   \, dc \\ \nonumber &=& \frac{1}{2}
e^{\frac{L}{8} (\alpha^2 + 6 \alpha -3)} \left \{ \mathrm{Erf}
\left[ \frac{\alpha^2 + 4 \alpha -1}{2(1+ \alpha)}
\sqrt{\frac{L}{2}} \right] +   \mathrm{Erf} \left[ \frac{1 -
\alpha}{2} \sqrt{\frac{L}{2}} \right] \right \}
\end{eqnarray}

Now, we focus on the regime $L \gg 1$ and,  according to the value
of the (finite) parameter $\alpha$, we distinguish among the
following cases:
\begin{itemize}
\item $\alpha = 1$\\
The expressions in Eqs.
(\ref{eq:p_alfa_L2}-\ref{eq:J_alfa_L}) can be evaluated exactly
obtaining, respectively:
\begin{equation}
p_{1,L} \approx \frac{1}{2} \mathrm{Erf} \left( \sqrt{\frac{L}{2}} \right)
= \frac{1}{2} \left[ 1 - \mathcal{O} \left( \frac{e^{-L/2}}{\sqrt{L}} \right) \right],
\end{equation}
\begin{equation}\label{eq:dp_da}
\frac{\partial p_{\alpha,L}}{\partial \alpha} \bigg|_{\alpha=1} \approx
 \frac{1}{2} \sqrt{\frac{L}{2 \pi}} ,
\end{equation}
and
\begin{equation}
\langle \tilde{J} \rangle_{1,L} \approx \frac{1}{2} \; e^{\frac{L}{2}} \;  \mathrm{Erf} \left(  \sqrt{\frac{L}{2}} \right ) = \frac{1}{2} e^{\frac{L}{2}} \left[ 1 - \mathcal{O} \left( \frac{e^{-\frac{- L}{2}}}{\sqrt{L} } \right) \right]  \approx e^{\frac{L}{2}} p_{\alpha,L}.
\end{equation}

\item $\alpha <1$ \\
\begin{equation} \label{eq:p_alfa_L_apicc}
p_{\alpha,L} \approx \sqrt{\frac{1}{2 \pi L}} \; \frac{1}{\tilde{\alpha}} \;
e^{-\frac{L}{2} {\tilde{\alpha}}^2} \left[ 1 + \mathcal{O} \left( \frac{1}{L} \right) \right],
\end{equation}
hence, $p_{\alpha,L} \to 0$ as $L \to \infty$. Moreover, for
$\alpha > - 1 + \sqrt{2} \approx 0.41$
\begin{equation} \label{eq:J_alfa_L_apicc}
\langle \tilde{J} \rangle_{\alpha,L} \approx  e^{\frac{L}{8}
(\alpha^2 + 6 \alpha -3)} \left[ 1 - \mathcal{O} \left( \frac{e^{
-\frac{L(1- \alpha)^2}{8} }}{\sqrt{L}} \right) \right],
\end{equation}
which is diverging for $\alpha > - 3 + 2 \sqrt{3} \approx 0.46$.

\item $\alpha > 1$\\
\begin{equation}
p_{\alpha,L} \approx 1 - \mathcal{O} \left(
\frac{e^{- \frac{L}{2} \tilde{\alpha}^2}}{\sqrt{L}} \right),
\end{equation}
hence $p_{\alpha,L} \to 1$ as $L \to \infty$. Moreover, analogously to the previous case,
\begin{equation} \label{eq:J_alfa_L_agran}
\langle \tilde{J} \rangle_{\alpha,L} \approx  e^{\frac{L}{8}
(\alpha^2 + 6 \alpha -3)} \left[ 1 - \mathcal{O} \left( \frac{e^{
-\frac{L(1- \alpha)^2}{8} }}{\sqrt{L}} \right) \right].
\end{equation}

\end{itemize}

\begin{figure}[tb]
\begin{center}
\resizebox{0.65\columnwidth}{!}{\includegraphics{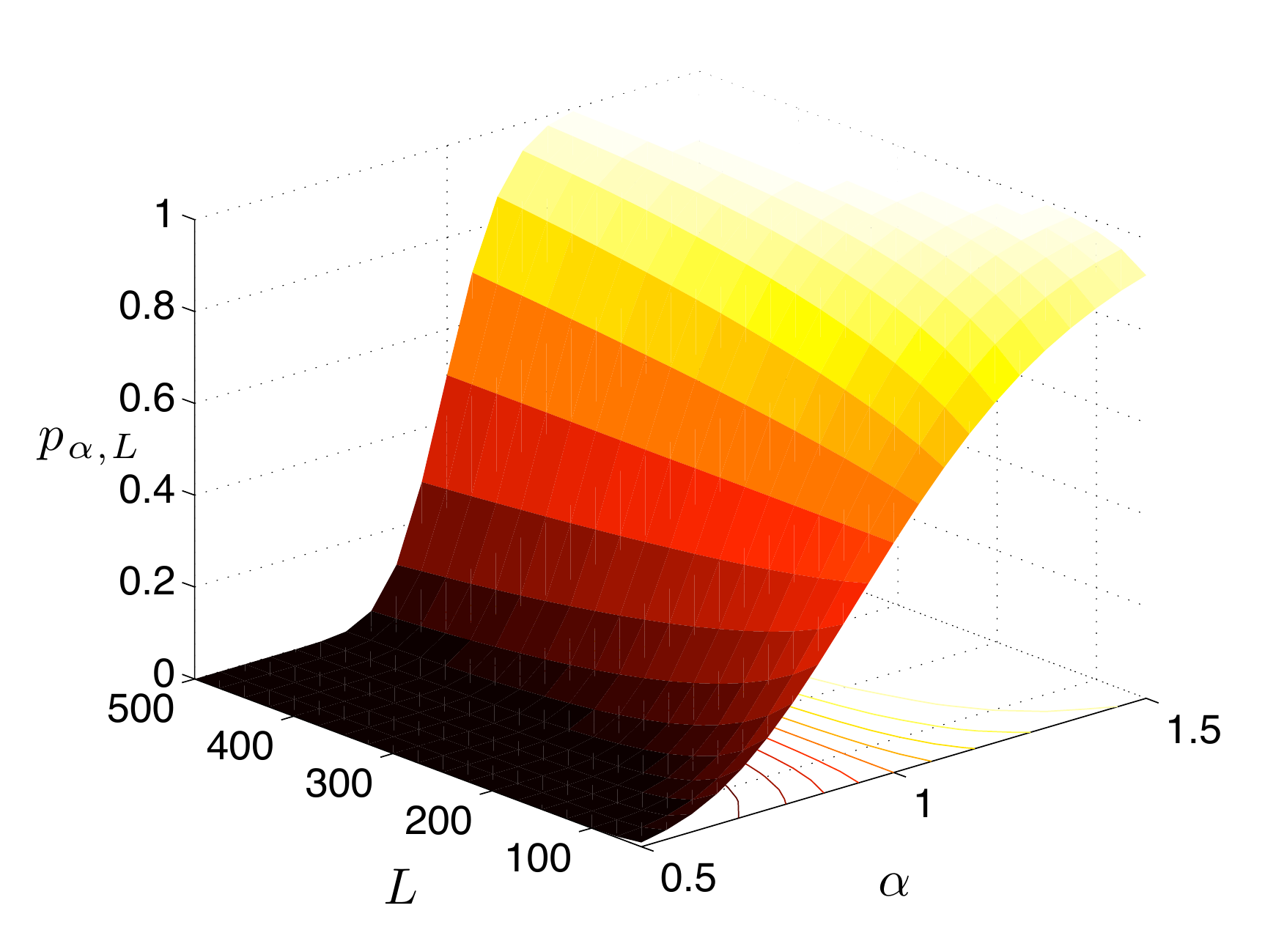}}
\caption{Link probability $p_{\alpha, L}$ for a system of $N=4000$ nodes as a function of $\alpha$ and of $L$. \label{fig:distr}}
\end{center}
\end{figure}

The asymptotic expressions above are all consistent with numerical
results which indeed confirm the validity of the Gaussian
approximation already for $L \sim 10^2$. Moreover, we notice that
in the limit $L \rightarrow \infty$ the link probability $p_{1,L}$
is a step function with diverging derivative in $\alpha = 1$ and
$p_{\alpha,L}=1$ for $\alpha>1$, while $p_{\alpha,L}=0$ for
$0 \leq \alpha <1$.
In Fig. (\ref{fig:distr}) we show the behavior of $p_{\alpha,L}$
as $L$ and $\alpha$ are varied.

In order to characterize the dilution of the graph under study, a
proper parameter is the average coordination number $\langle k
\rangle = p_{\alpha,L} N $, as it represents the average number of
links stemming from a node. In principle, $\langle k \rangle$
depends on $\alpha, L$ and $N$, which in our model cover a well
precise physical and biological meaning. In fact, while $\alpha$
derives from the chemical-physical interactions arising among
antibodies and antigens, and can be set independently of $N$ and
$L$, the latter parameters $N$ and $L$ are intrinsically connected
with each other. First of all, for strings of length $L$, the
minimum number $N$ of agents has to be lower than $2^L$, if we
want to avoid repetitions. Actually, as underlined in Sec.
(\ref{sec:1B2B}), $N$ is expected to be much smaller than $2^L$.
Moreover, a proper scaling of the system size should preserve its
(global) topological features, namely $\langle k \rangle$. In
particular, $\langle k \rangle$ must be finite in order to have a
well-defined thermodynamic limit for $N \to \infty$
\cite{a13,a52,a90}; all other cases would be either trivial
($\langle k \rangle \to 0$) or un-physical ($\langle k \rangle \to
\infty$). Thus, on the one hand $\langle k \rangle$ specifies the degree of
dilution of our idiotypic network, on the other hand it crucially affects the
statistical mechanics of the whole system. Therefore, it is natural to describe a system by means of the three parameters $\alpha$,  $L$ and $\langle k \rangle$. Then, the number of nodes $N$ follows as\begin{equation} \label{eq:scaling}
N = \frac{\langle k \rangle}{p_{\alpha,L}} \approx  \frac{2 \langle k \rangle}{\mathrm{Erf}\left( \sqrt{\frac{L}{2}} \right) - \mathrm{Erf}\left( \tilde{\alpha} \sqrt{\frac{L}{2}} \right) },
\end{equation}
where we used Eq.~(\ref{eq:p_alfa_L2}). In particular, for $\alpha
<1$, one can write (see Eq.~\ref{eq:p_alfa_L_apicc})
\begin{equation} \label{eq:scaling2}
N \approx \sqrt{2 \pi L} \; \langle k \rangle  \tilde{\alpha} \; e^{\frac{L}{2}
\tilde{\alpha}^2},
\end{equation}
which should be compared with Eq. (\ref{NL}), to get $f(L) \sim
\sqrt{L}$.

\begin{figure}[tb]
\resizebox{7.0cm}{6.5cm}{\includegraphics{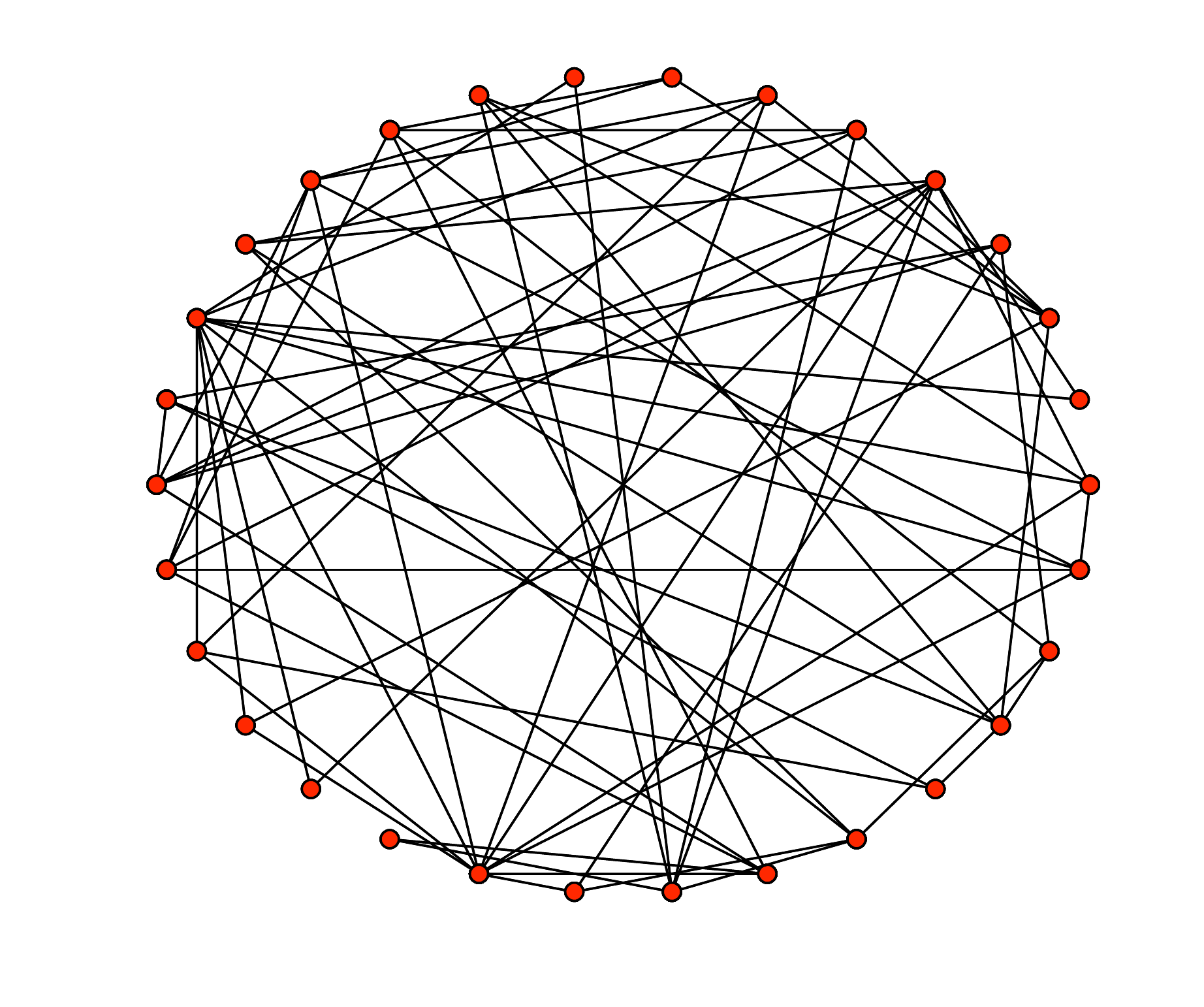}} \label{fig:ER}
\resizebox{7.5cm}{9.0cm}{\includegraphics{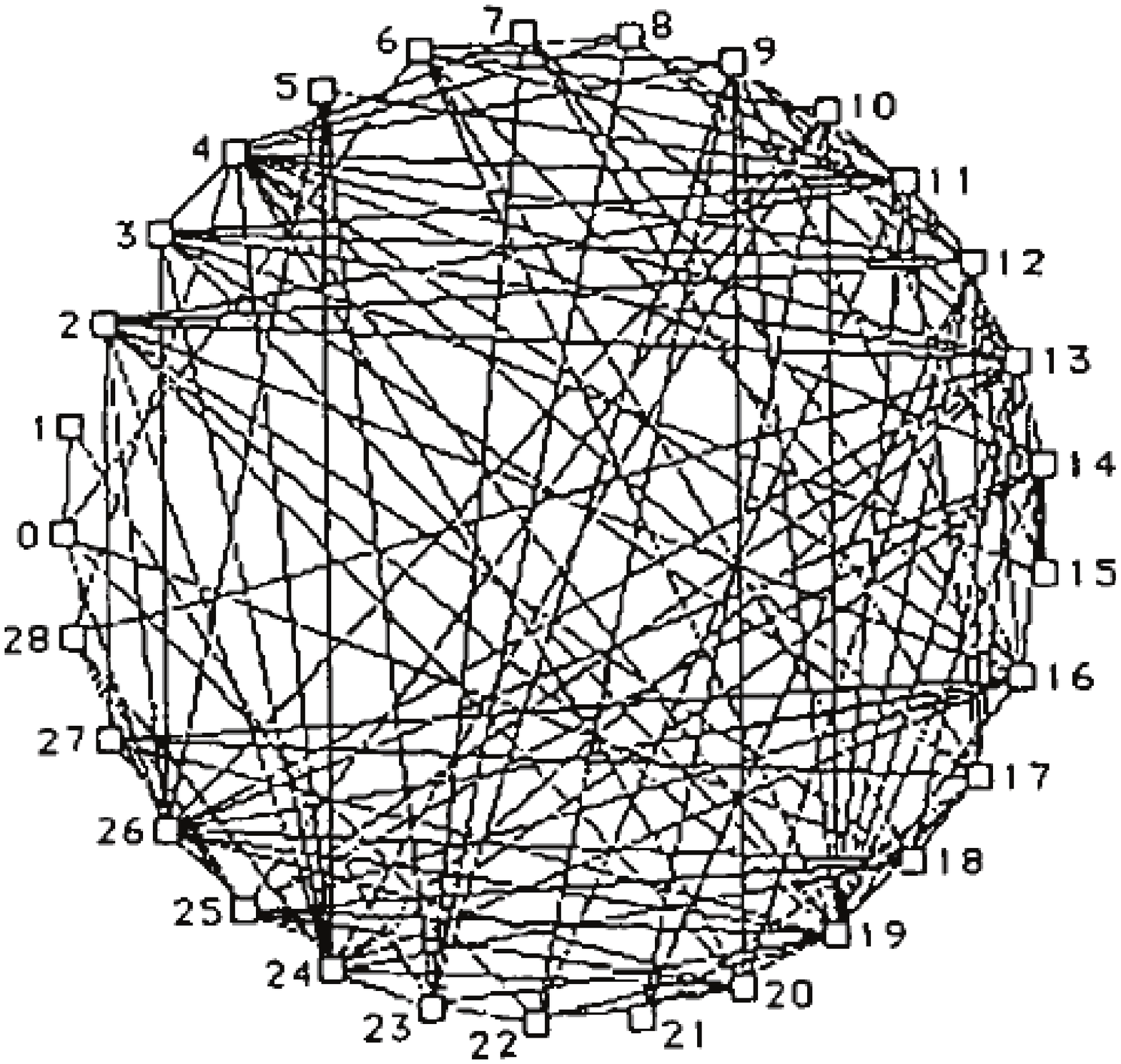}} \label{fig:Varela}
\caption{Left panel: Idiotypic network obtained for $\alpha=0.7$, $L=140$; for the sake of clarity we fixed a small number of nodes $N=30$. Right panel: Representation of the connectivities among a group of neonatal antibodies found by Varela and Coutinho in \cite{a39}.}
\end{figure}

Finally, we notice that several independent experiments, lead on
natural antibodies in neonatal mice, have evidenced that the
network is highly connected: each idiotype is able to recognize on
average $20-25 \%$ of a given panel
\cite{a106,bagley,farmer}. However, such connectivity
decreases with time and it is expected to reach a steady state value, as
corroborated by higher values of dilution in adult mice \cite{holmberg}; in particular, in \cite{per_string} the connectivity in adult immune systems is estimated around $3 - 5 \%$. Following these data, in our
system, meant for a mature immune system, the link probability
$p_{\alpha,L}$ is expected to be approximately $0.04$, which
provides a first hint for selecting the region $\alpha<1$.
Moreover, recalling $N \sim \mathcal{O}(10^{14})$, we consistently expect
$\langle k \rangle \sim \mathcal{O}(10^{12})$.
Therefore, in our model a realistic system may be obtained by fixing $\alpha = 0.7$, $L=140$ and $\langle k \rangle \sim \mathcal{O}(10^{12})$,
then from Eq. (\ref{eq:p_alfa_L}) we get $p_{\alpha,L} \approx
0.04$, from which we recover $N
\sim \mathcal{O}(10^{14})$; the whole framework
is therefore quantitatively consistent with real data.

We conclude this section by showing a typical topological structure corresponding to parameters $\alpha = 0.7$ and $L=80$, see Fig.~\ref{fig:ER}, left panel; a comparison with the network obtained experimentally in \cite{a39}, right panel, is also provided. The similarity between the two structures is manifest and suggests that the Erd\"{o}s-Renyi topology which emerges from our minimal assumption is consistent with real data.

\section{Weighted connectivity} \label{sec:weight}
In Sec. (\ref{sec:graphs}) we introduced the degree $k_i$ which
represents the number of lymphocytes ``in contact'' with the
lymphocyte labeled as $i$ and making up the set denoted as $V_i
\subseteq V$. This means that the lymphocyte $i$, can
interact with lymphocytes in $V_i$ through the pertaining
antibodies described by the string $\xi_i$.

The coordination number has been shown to play a crucial role in
the reactivity of antibodies \cite{a38}: the patterns of
cross-reactivity between collections of idiotypes have been
organized by experimental immunologists in matrices which are just
the adjacency matrices describing the affinity among idiotypes.
Such matrices have been revealed to be be arranged in blocks: a
high-connectivity block (including nodes characterized by large
degree), a ``mirror block'' (intermediate degree) and a
low-connectivity block (small degree). These blocks were analyzed
in their independent contributions through simulations by Varela
et al. \cite{a38} who concluded that the larger the degree, the
lower the reactivity of the corresponding lymphocytes and the
greater their degree of tolerance. Hence the various groups play
different roles:  the mirror group accounts for intrinsic
oscillations, while highly connected nodes may act as initial
organizers of the immune system. In this framework, and according
to the autopoietic view, autoimmunity arises not because of the
presence of self-reactive clones, but because such clones
are or become  ``not properly'' connected to the network. As a
result, self/non-self discrimination turns out to be an emergent
property of the immune network which is therefore able to organize
the mature repertoire. It follows naturally that establishing the
structure of the immune system is a task of great importance as it
would allow to figure out possible strategies to cope with
autoimmune diseases.

In our model the adjacency matrix is actually weighted since links
are endowed with a weight $J_{ij}$, which allows to introduce a
weighted degree $w_i$ as
\begin{equation}\label{eq:connectivity}
w_i(\alpha,L,\langle k \rangle) \equiv \sum_{j=1}^N J_{ij} (\alpha,L).
\end{equation}

Notice that the local quantity $w_i$ provides finer information
with respect to $k_i$, being directly connected with the
"internal" stimulus felt by lymphocyte $i$: recalling the
Hamiltonian of Eq. (\ref{eq:H_2}), and assuming, for the sake of simplicity, the zero noise limit so that all
lymphocyte in $V_i$ are quiescent ($m_j = -1, \forall j \in V_i$), the local field acting on
$i$ is just $\varphi_i = -\sum_{j=1}^N J_{ij} m_j = w_i$.

For a given system $(\alpha, L, \langle k \rangle)$ the average weighted degree can be calculated as
\begin{equation}
\langle w \rangle_{\alpha,L,\langle k \rangle} =  \sum_k P(k) \sum_{c_1, c_2, ... , c_k=L/(\alpha+1)}^{L} \prod_{i=1}^k \mathcal{P}(c_i) \sum_{i=1}^k \frac{e^{c_i(\alpha+1)-L}}{\sum_{c_i} \mathcal{P}(c_i) e^{c_i(\alpha+1)-L}}.
\end{equation}
Actually, it is convenient to introduce the "quenched averages" $\bar{J}(\alpha,L)$ and $\bar{w}(\alpha,L,\langle k \rangle)$, obtained by averaging the couplings $J_{ij}$ and the weighted degree $w_i$ over all links and nodes, respectively, of a given realization, namley
\begin{equation} \label{eq:av_w}
\bar{J} =   \frac{\sum_{i=1}^N \sum_{j=1}^N J_{ij}}{N(N-1)} =  \bar{J},
\end{equation}
and
\begin{equation} \label{eq:av_w}
\bar{w} = \frac{\sum_{i=1}^N w_i}{N} =  \frac{\sum_{i=1}^N \sum_{j=1}^N J_{ij}}{N} = (N-1) \bar{J}.
\end{equation}
Now, for large $N$ and $L$ the "quenched averages" $\bar{J}(\alpha,L)$ and $\bar{w}(\alpha,L,\langle k \rangle)$ converge to the "ensemble averages" $\langle J
\rangle_{\alpha,L}$, that is
$\langle w \rangle_{\alpha,L,\langle k \rangle} \approx (N-1) \langle J \rangle_{\alpha,L}$ and, being that
$\langle J \rangle_{\alpha,L}$ equals one by definition, we have that $\langle w \rangle_{\alpha,L,\langle k \rangle}$ scales linearly with $N$.
Analogously, due to the
uncorrelatedness among $J_{ij}$'s, one can use Bienayme
 prescription and write
\begin{equation}
\mathrm{Var} (w_i) = \mathrm{Var} \left( \sum_{j=1}^N J_{ij} \right) = \sum_{j=1}^N \mathrm{Var} (J_{ij}) = N \mathrm{Var} (J_{ij})
\end{equation}
where $\mathrm{Var}(x) \equiv \bar{x^2} - \bar{x}^2 \equiv
\sigma_{x}^2$ is the variance of the variable $x$, that is the
expected value of the square of the deviation of $x$ from its own
mean $\bar{x}$. Now, the variance for $J_{ij}$ can be estimated
via Eq. (\ref{eq:J}) and Eq.(\ref{eq:P_c})
\begin{eqnarray} \label{eq:J2_alfa_L}
\nonumber
\langle J^2 \rangle_{\alpha,L} &=&
\frac{1}{\langle \tilde{J} \rangle_{\alpha,L}^2} \int_{\frac{L}{\alpha+1}}^L \exp{\left[2 c(\alpha + 1) - 2L \right]}
\sqrt{\frac{2}{\pi L}} e^{- \frac{(c - L/2)^2}{L/2} } \; dc \\
\nonumber
&=& \frac{e^{\frac{L}{2}(\alpha^2 + 4 \alpha -1)}}{2\langle \tilde{J} \rangle_{\alpha,L}^2}
 \left[ \mathrm{Erf} \left( \frac{\alpha(3+\alpha)}{1 + \alpha} \sqrt{\frac{L}{2}} \right) - \mathrm{Erf} \left( \alpha \sqrt{\frac{L}{2}} \right) \right].
\end{eqnarray}
Now, with some algebra and recalling Eq.~(\ref{eq:J_alfa_L}), we get the following estimate
\begin{eqnarray}\nonumber
\langle J^2 \rangle_{\alpha,L}  &\approx&
2 e^{\frac{L}{4}(\alpha+1)^2 }  \\ \label{eq:J2_ex}
&& \times \frac{ \left[ \mathrm{Erf} \left( \frac{\alpha(3+\alpha)}{1 + \alpha} \sqrt{\frac{L}{2}} \right) - \mathrm{Erf} \left( \alpha \sqrt{\frac{L}{2}} \right) \right] }{ \left[ \mathrm{Erf}
\left(\frac{\alpha^2 + 4 \alpha -1}{2(1+ \alpha)}
\sqrt{\frac{L}{2}} \right) +   \mathrm{Erf}
\left( \frac{1 - \alpha}{2} \sqrt{\frac{L}{2}} \right) \right]^2 }   \\
&=& \frac{1}{2} e^{\frac{L}{4}(- \alpha^2 + 2 \alpha +1)} \frac{1}{\alpha \; \sqrt{2 \pi L}} \left[ 1 - \mathcal{O} \left (\frac{1}{L} \right) \right].
\end{eqnarray}
After noticing that, for $1 - \sqrt{2} < \alpha < 1 + \sqrt{2}$ the exponent is positive, yielding $\langle J^2 \rangle_{\alpha,L} \gg \langle J \rangle_{\alpha,L}^2 = 1$, we can write the standard deviation for the coupling strength as
\begin{equation}
\sigma^J_{\alpha,L} \approx  \sqrt{ \langle J^2 \rangle_{\alpha,L}}  \sim
\frac{1}{\sqrt[4]{L}} e^{\frac{L}{8}(- \alpha^2 + 2 \alpha +1)}.
\end{equation}
Moreover, one can write
\begin{equation}
\sigma^w_{\alpha,L} \approx \sqrt{ N  \langle J^2 \rangle_{\alpha,L}} \approx \sqrt[4]{L}\; e^{\frac{L}{4} \frac{(-2 a^4 + 9 a^2 + 6a +3)}{(1+a)^2} },
\end{equation}
where in the last expression we used Eq. (\ref{eq:scaling2}).
%

\begin{figure}
\begin{center}
\resizebox{0.6\columnwidth}{!}{\includegraphics{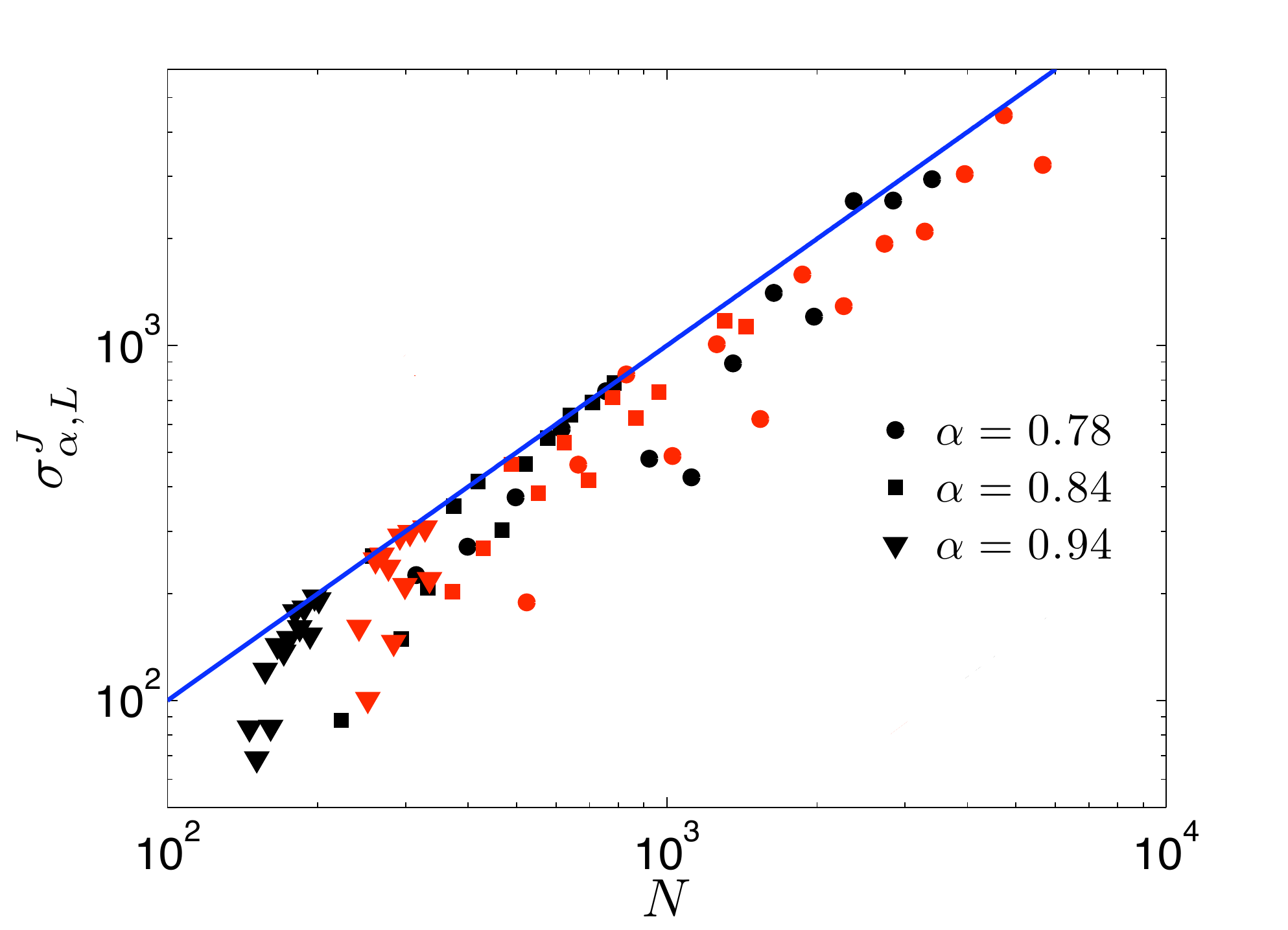}}
\caption{Standard deviation $\sigma^J_{\alpha,L}$ as a function of the system size $N$; the parameter $L$ is properly rescaled in order to keep $\langle k \rangle$ fixed and equal to $60$ (red) or to $100$ (black); different symbols represent different choises for $\alpha$, as shown by the legend. The straight line represents the first bisector.} \label{fig:variance}
\end{center}
\end{figure}

In Fig. (\ref{fig:variance}) we show, as a function of $N$ and
for several choices of $\alpha$, the standard deviation
$\sigma^J_{\alpha,L}$; notice that while $N$ is varied, $L$ is properly scaled in order to keep $\langle k
\rangle$ fixed. The log-log scale plot highlights a regime, for large enough $N$, where a power law
growth for $\sigma^J_{\alpha,L}$ holds.

\section{Circuits}
In this section we want to analyze the number and the relative
weight of small loops present in the random graph
$\mathcal{G}(N,p_{\alpha,L})$, previously introduced to describe
the idiotypic network. This kind of information allows not only to
deepen the topological description of the immune network, but it
will also be useful in the following sections when studying the
system response to external stimuli.

In fact, we recall that due to the lacking of a perfect match
among antibodies, a given lymphocyte, say $\sigma_1$, undergoing
clonal expansion, may elicit one (or more) of the best Jerne
counterparts (even spurious state may respond), say $\sigma_{2}$.
The latter undergoing clonal expansion too, may elicit another
lymphocyte among the best Jerne spurious state, say $\sigma_3$,
and so on. Now, since $\sigma_1$ and $\sigma_3$ both have large affinity, i.e. complementarity, with the same state $\sigma_2$, they are expected to be similar. Analogously, $\sigma_4$ is expected to be similar to $\sigma_2$ and therefore to disply a large affinity with both $\sigma_1$ and $\sigma_3$. As a result, such loops built by four lymphocytes which are mutually Jerne states are expected to be more
stable than the loops built by three lymphocytes.
\newline
Of course, again due to the multi-attachment (able to generate
spurious states) the information gets lost when increasing the
size of the loop such that large loops are unexpected; in the
following particular attention will be paid to loops of length $3$ and $4$.

Let us now formalize and analyze mathematically the problem. First of all, we define a circuit of length $l$ as the edge set
$\ell_l$ of an undirected closed path without repeated
edges, that is $\ell_l = \{ (i_1, i_2), (i_2, i_3), ...,(i_l, i_1) \},$
with $(i_k, i_{\mathrm{mod}(k+1,l)}) \in \Gamma$, for any $k \in
[1,l]$.

Then, we denote with $n_l$ the number of circuits of length $l$ present in the graph; it's worth recalling that $n_l$ is a purely topological quantity depending only on the adjacency matrix $\mathbf{A}$. Now, the number of possible circuits reads as $\frac{1}{2l} \frac{N!}{(N-l)!}$, in fact, choosing such a circuit implies to select an ordered list of $l$ vertices, modulo the orientation and the starting point of the path. Now, for an Erd\"{o}s-Renyi random graph the probability for a circuit to be effectively present in the network only depends on its length, being given by the link probability to power $l$; in particular, here we get
\begin{equation}
n_l= \frac{1}{2l} \frac{N!}{(N-l)!} p_{\alpha,L}^l.
\end{equation}
For short cycles with $l \ll N$, we can approximate the previous expression as
$$
n_l \approx \frac{1}{2l} \langle k \rangle^l,
$$
where we used that the
average degree in $\mathcal{G}$ is $\langle k \rangle = p_{\alpha,L}(N-1)$. Therefore, above the percolation
threshold, the number of circuits grows exponentially fast with
their length.

It is also possible to calculate the number of circuits of length $l$, such that they include a node of degree $k$; this is simply given by
\begin{equation}
n_l(k) = \frac{1}{2l} \frac{(N-3)!}{(N-l)!} p^{l-2} k(k-1) \approx \frac{k^2}{2lN} \langle k \rangle ^{l-2} ,
\end{equation}
where the last expression holds for $l$ small and $k$ large.  As a
consequence, as $l$ is increased, the number of circuits increases
exponentially and the rate of growth increases with $k$.

As underlined before, our network can not be described in a purely topological way, i.e. in terms of the adjacency matrix only, as the coupling strength associated to each link has an important physical meaning. Consequently, circuits should also be measured according to the couplings $J_{ij}$, or weights $w_i$, of the pertaining links or sites, respectively: the overall strength $J_{\ell_l }$ associated to the circuit denoted as $\ell_l = \{ (i_1, i_2), (i_2, i_3), ...,(i_l, i_1) \}$, is therefore given by
\begin{equation}
J_{\ell_l} = \frac{1}{l} \sum_{k=1}^l J_{i_k,i_{k+1}},
\end{equation}
being $i_{l+1} \equiv i_1$, while the overall weight is
\begin{equation}
w_{\ell_l} = \frac{1}{l} \sum_{k=1}^l w_{i_k}.
\end{equation}
Notice that $J_{\ell_l}$ represents the intrinsic robustness of the circuit $\ell_l$, while $w_{\ell_l}$ represents the overall (relative) influence from the external environment to the circuit $\ell_l$.
In the following we highlight the role of the quantities $J_{\ell_l}$ and of $w_{\ell_l}$, by considering a special situation.

Let us assume, for the sake of simplicity, that we are in the
zero noise limit, so to discard the $\beta$ dependence of the
quiescent state. In this condition, and in the absence of external stimuli we have $m_k = -1, \forall k \in V$. Now, when a sufficiently high concentration of the antigen $\bar{\xi}_i$ is introduced, the $i$th clone will undergo a clonal expansion, so that $m_i>-1$.

Then, the stimulus can spread from $i$ to nodes in
$V_i$ and so on throughout the network in a cascade fashion,
possibly coming back to $i$, hence providing a reinforcement
feedback. In this case we say that a ``firing circuit" has
established. Given a circuit $\ell_l$ we say that it is firing if $m_{i_k} > -1$, for any
$k \in [1,l]$. The magnitude of the firing circuit can be measured
in terms of global firing concentrations, namely $\sum_{k=1}^l c_{i_k}/l \in
[1,M]$, see Eq.~(\ref{eq:concetration}).

Due to
self-reinforcement, a firing circuit can survive even when the external stimulus has expired; the long-time persistence of a firing circuit can be estimated by means of the ratio $J_{\ell} / w_{\ell}$: the larger the ratio, the more important the self-reinforcement with respect to the neighborhood (possibly non-firing) and therefore the more likely the persistence.
Interestingly, the persistence of firing circuits yields
storage memory of previous infection, hence, finding the conditions for their establishment can be very important.

\begin{figure}[tb]
\resizebox{0.8\columnwidth}{!}{\includegraphics{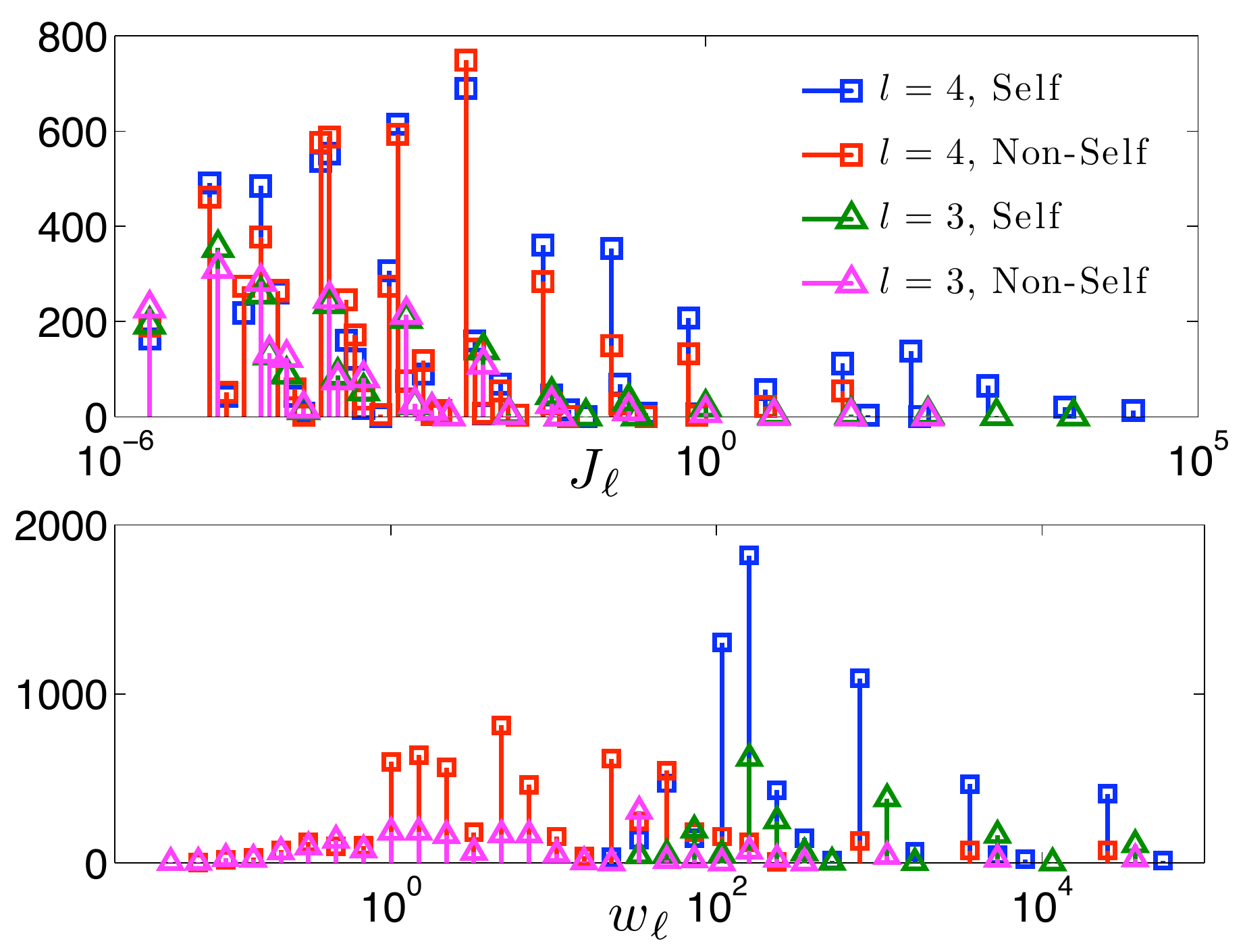}}
\caption{Distributions for overall strength $J_{\ell}$ (upper panel) and overall weight (lower panel) concerning circuits of length $l=3$ (triangles) and $l=4$ (squares). We also distinguished between circuits passing through lowly-conned nodes (Non-self, represented in red and in pink) and through highly-connected nodes (Self, represented in blue and in green), as shown by the legend.
From the upper panel it is clear that for $l=4$ the overall strength is larger than for $l=3$.
The graph considered is made up of $N=500$ nodes, moreover $\alpha=0.74$ and $L=80$.
\label{fig:circuiti} }
\end{figure}

In this context we are
interested in short circuits rather than long $\mathcal{O}(N)$ ones, the latter
evidently require more strict conditions for their establishment and should actually be avoided as they would correspond to a broad response involving most lymphocytes,
including those directed against self.
In particular, we focus the attention on circuits of length $l=3$ and $l=4$, which are expected to  result in sensitively different overall strengths $J_{\ell}$. In fact, the former corresponds to a frustrated configuration giving rise to a relatively small overall strength. More precisely, since links derive from a large degree of complementarity between the two adjacent nodes, circuits of odd, small length must display at least one weak edge; on the other hand, when the length is even this kind of frustration can be avoided.

As for the overall weight $w_{\ell}$, its value strongly depends on whether the circuit contains a highly-connected node, namely a node belonging to the left-hand side of the distribution $P(w)$; indeed, as we will see in the following section, such nodes are typically connected with analogously highly-connected nodes.

In Fig.~\ref{fig:circuiti} we show the distributions obtained for
$J_{\ell}$ (upper panel) and for $w_{\ell}$ (lower panel), when
considering all the circuits passing through lowly-connected and
highly-connected nodes respectively; both cases $l=3$ and $l=4$
are considered. From the data depicted  in the figure we
calculated the average overall strength which, for $l=4$ turns out
to be more than twice the one pertaining to $l=3$; the former is larger than $1$, the latter is smaller than $1$ (we recall that $1$ is the expected coupling strength). In fact, as anticipated, a lymphocyte $\sigma_1$ and its corresponding anti-lymphocyte are unlikely to have a strong common neighbour.
We also notice that the presence of a highly connected node within the circuit has dramatic effects on the distribution for $w_{\ell}$ which results to be shifted towards larger values; this means that circuits involving self-addressed nodes (i.e. those with high connectivity, vide infra) are also more unlikely to be elicited.

\chapter{Features of the model}

\section{Self/Non-Self recognition} \label{sec:recognition}
Let us consider again the whole idyotipic network made of $N$
different clones, each characterized by a specific string of $L$
idyotopes; once $\alpha$ is fixed, the affinity between two
different nodes is specified by Eq. (\ref{eq:affinity}) from which
the coupling in Eq. (\ref{eq:J}) follows.

Before turning to the analysis of the distribution $P(w)$ and showing how it naturally allows to distinguish between self and non-self addressed antibodies, it is worth recalling the famous experiment lead by Stewart, Varela and Coutinho \cite{a38,a39}: they measured the affinity of a collection of antibodies and analyzed the related affinity matrices finding that these matrices are organized in blocks. More precisely, they distinguished a high-affinity block, two blocks of groups which are mirror of each other, and a low affinity remnant; then they showed that various groups play different roles: the mirror groups provide their model with various oscillations periods, while highly connected nodes maintain a ``basic network background level" and may be looked at as self-addressed antibodies. Indeed, their large connectivity prevents them from readily react to a stimulus. This point of view is extremely interesting as it outlines a natural interpretation of the topological properties of the immune network. Not only, from an autopoietic point of view, it also sheds light on autoimmunity diseases: their origin would therefore lay on the ``inadequate connection'' of self-reactive clones.

Let us now consider the idiotypic network introduced in the previous section: with respect to the model introduced by Coutinho and Varela \cite{a39}, here each link stemming from a given node $i$ has its own weight, which measures how strong the affinity between the relevant antibodies is, in such a way that, the information carried by $w_i$ is much reacher than the one carries by $k_i$. Hence, we will follow
the experiment by Stewart, Varela and Coutinho
\cite{a38,a39} by looking \emph{not} at the degree distribution $P(k)$, but rather at the distribution $P(w)$.

\begin{figure}
\begin{center}
\resizebox{0.45\columnwidth}{!}{\includegraphics{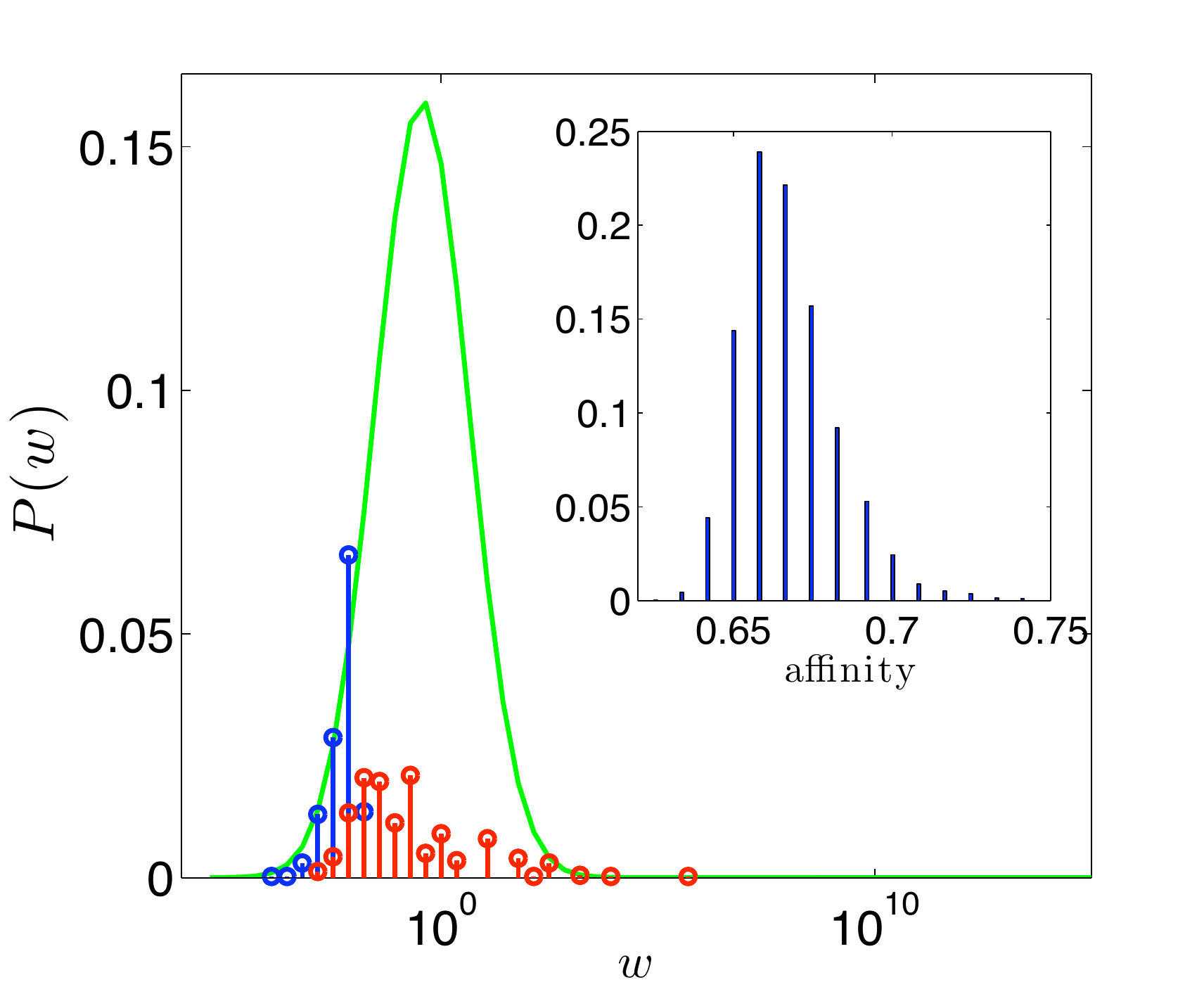}}
\resizebox{0.45\columnwidth}{!}{\includegraphics{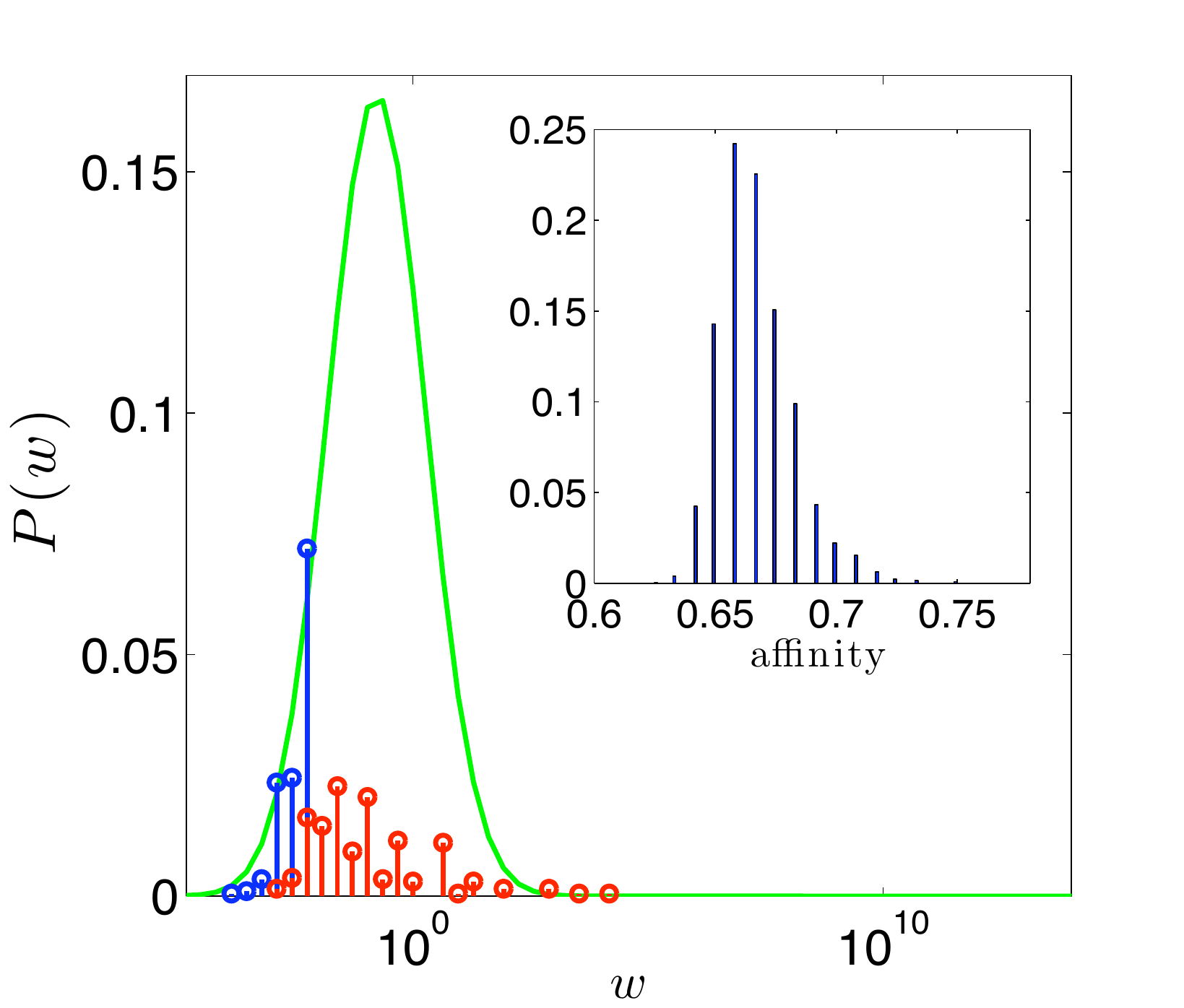}}
\resizebox{0.45\columnwidth}{!}{\includegraphics{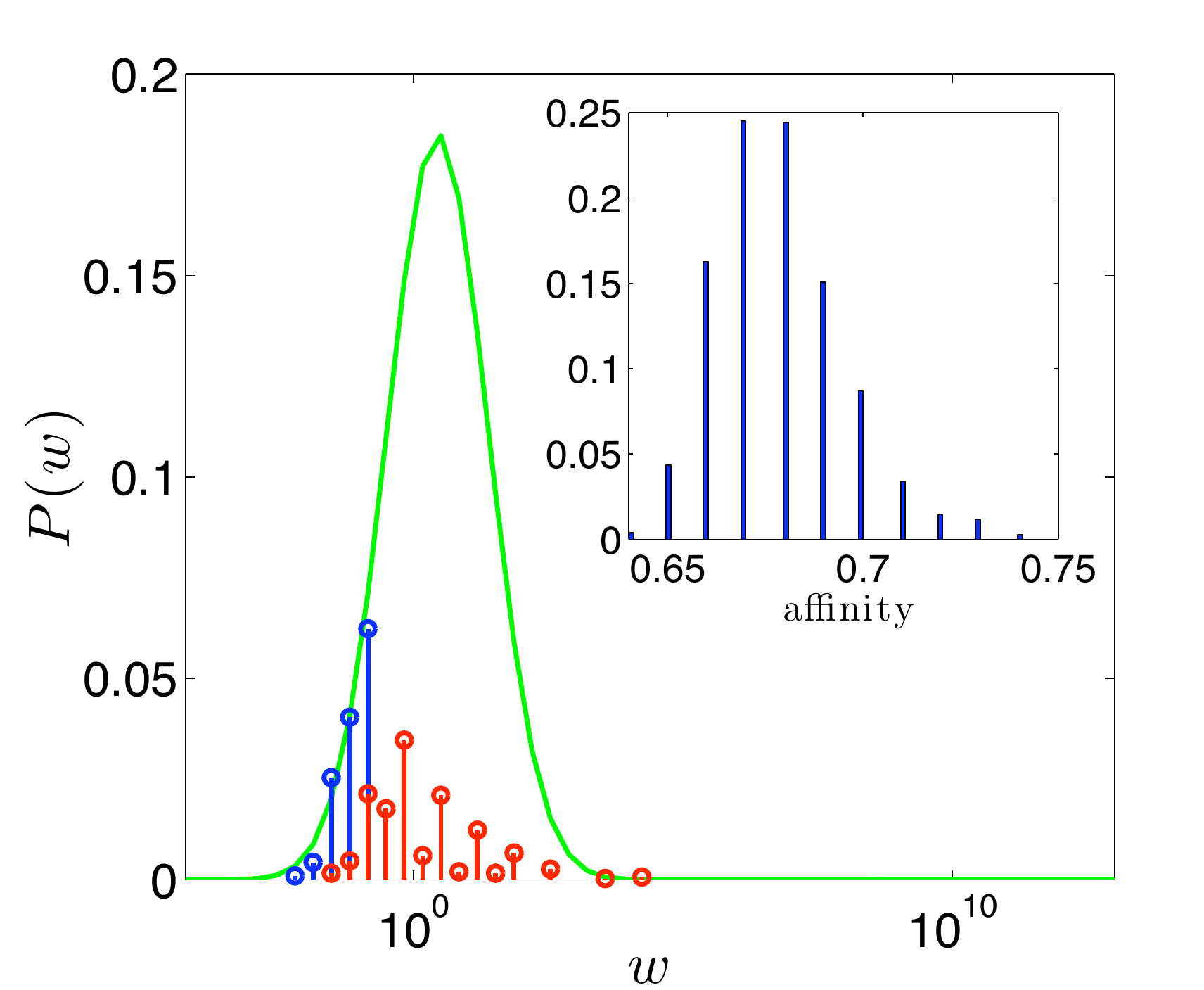}}
\resizebox{0.45\columnwidth}{!}{\includegraphics{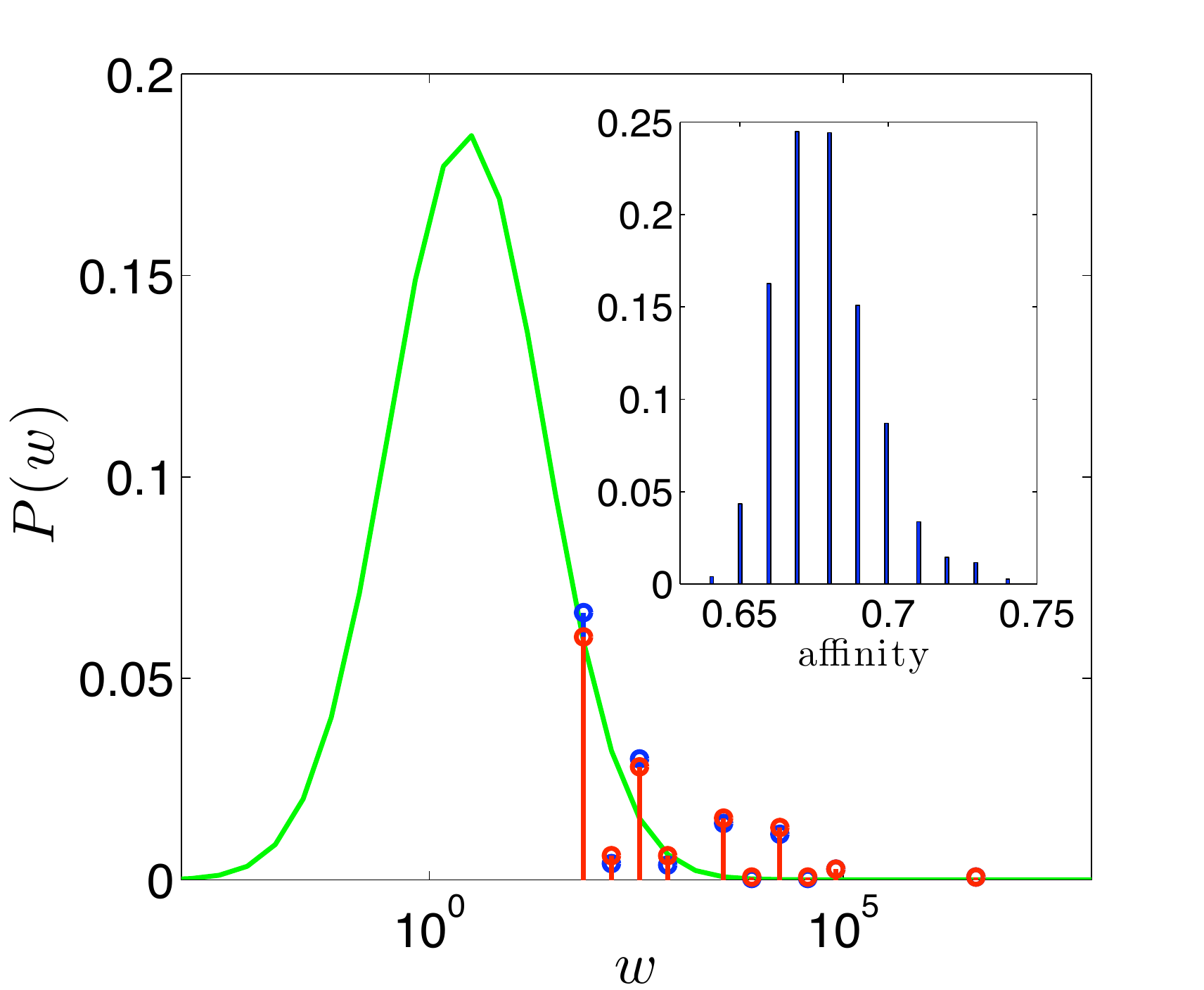}}
\caption{Weighted connectivity distributions (green line) for systems made up of $N=4000$ lymphocytes characterized by idiotypic strings of length $L=120$ and $\alpha=0.74$ (panel a), $L=120$ and $\alpha=0.82$ (panel b), $L=100$ and $\alpha=0.78$ (panel c,d); the average connectivity $\langle k \rangle$ gets $240$, $630$ and $290$, respectively. In blue we show the distribution of low-connected clones (panels $a$, $b$ and $c$) or high-connected clones (panel $d$) chosen and in red the distribution of the pertaining "anti-clones". Notice the semilogarithmic scale plot. The insets show the relevant matching $c_{ij}$ among all the couples Ig and anti-Ig detected in the whole system. \label{fig:weight} }
\end{center}
\end{figure}

We considered different systems $(\alpha,L,\langle k \rangle)$ and by numerical analysis we derived the distribution $P(w)$, which, on a semilogaritmic scale, can be properly fitted by a Gaussian distribution; the relevant best fits are represented by the green curves in Fig.~\ref{fig:weight}. Such distributions naturally outline three main groups
characterized by high (right-hand-side tail), intermediate (central region) and low (left-hand-side tail) weighted degree,
respectively. Hence, recalling that a large weight implies a low reactiveness (see previous section), lymphocytes displaying low and high weighted degree can be labelled as non-self and self addressed lymphocytes, respectively. It is important to notice that, since the distribution $P(w)$ covers several orders of magnitude,
the former will easily react even by
low dose of nearest-neighbors (or antigens) (implicitly defining
the low dose tolerance), while, for larger $w$, the ability to
react decreases progressively, up to prohibitive values of antigenic concentration.

Now, starting from $P(w)$ we want to focus on couples of Ig and anti-Ig and figure out possible correlations in their weighted degrees.
We first select non-self addressed lymphocytes,
namely modes in the network corresponding to the left-hand side of the weighted degree
distribution, and we look for their most tightly connected
neighbors among the remaining $N-1$. This way, we distinguish
couples $(i, \bar{i})$, where $i$ should be meant as a lymphocyte
producing Ig directed against non-self agents and $\bar{i}$ as a
lymphocyte producing so called anti-Ig able to respond to a
significant growth in Ig's concentration, according to Jerne's idea
of idiotypic network.

As shown in Fig. (\ref{fig:weight}), when $i$ belongs to the low
weighted-degree region, the corresponding $\bar{i}$ typically
falls in the intermediate region of the distribution, hence
fitting the ``mirror block''; this holds for several choices of
$\alpha$, $L$ and $\langle k \rangle$ (see panels $a$, $b$ and $c$).

Let us now turn to the right-hand side of the weighted-degree
distribution and, analogously, we distinguish couples $(j,
\bar{j})$, where $j$ represents a lymphocyte producing Ig directed
against self agents and $\bar{j}$ as the relevant anti-Ig able
(see panel $d$). In this case anti-Ig still belongs to the
highly-connected group, that is they should as well be meant as
directed to self-agents. This result is easy to see: Since by
definition $j$ exhibits a large weight $w_j$, it follows that,
typically, $J_{j \bar{j}} \sim w_j$ as the main contribution to
$w_j$ comes from $J_{j \bar{j}}$ and, analogously, $J_{j \bar{j}}
\sim w_{\bar{j}}$. Otherwise stated, when a highly connected node
is selected, there exists a correlation between $w_j$ and
$w_{\bar{j}}$; on the other hand when a lowly connected node is
considered, the contribution of $J_{i \bar{i}}$ to $w_i$ and to
$w_{\bar{i}}$, respectively, is small enough not to bias
$w_{\bar{i}}$. The
very origin of such a different behavior of self and non-self
anti-Ig lays in the wide range spanned by $w$.

As deepened in the following,
anti-Ig's play a crucial role in the establishment of memory
effects, so that the ``mirror block'' here acquires the
fundamental function of memory storage. Interestingly, such a memory storage here turns out to be effectively managed since it is restricted to non-self directed Ig only. Conversely, self directed Ig and relevant anti-Ig both set up the highly-connected group.

Another point worth being underlined concerns the affinity between
Ig and anti-Ig; the affinity can be evaluated from the number of
matchings $c_{ij}$, whose distribution is represented in the
insets of Fig. (\ref{fig:weight}): due to the uniform distribution
underlying the extraction of $\xi$'s, the relative matching
$c_{ij}/L$ resulting from all possible pairs is distributed around
the value $1/2$ with a standard deviation scaling as $1/\sqrt{L}$.
Hence, the expected affinity for Ig and anti-Ig can be estimated
as $1/2(1 + 3/\sqrt{L})$. Recalling that in our model each entry
in $\xi$ represents an epitope and that each Ig displays
$\mathcal{O}(10^2)$ epitopes, we have that the relative matching
between Ig and anti-Ig falls in between $0.6$ and $0.7$. It is
interesting to stress that experimental values are bounded by
$0.6$ (and in general larger): This can be understood by analyzing
a $l=4$ Jerne loop and measuring the binding ability of the last
antibody Ab4 with the first Ab1. On average $>60\%$ of Ab4
molecules are able to link Ab1 \cite{a75}.

The consistency among such data confirms the plausibility of
interpreting $\xi$ as a string of $\mathcal{O}(10^2)$ epitopes (even thought
improvements in the binding percentage should be achieved by using
more complex epitope distribution probabilities). Also, the
relative small affinity found experimentally among most-tightly
nodes corroborates the fact that the system is far from complete,
namely that $N \ll 2^L$, consistently with the assumption of
Eq.~(\ref{NL}).

\section{Low-dose tolerance}\label{sec:lowdose}
In this section we want to investigate the effects elicited by a
concentration $c$ of a given antigen. Let us consider the antigen
with specificity $\xi_{h^i} \equiv \bar{\xi_i}$, namely displaying
perfect match with antibody $\xi_i$. The presence of a
concentration $c$ of the antigen can be incorporated within the
Hamiltonian $H$ describing the system by introducing an external
field $\mathbf{h^i}$, whose element $h^i_k$ represents the
coupling with the $k$-th antibody, namely
\begin{equation}
h^i_k = \exp{[f_{\alpha,L}(\xi_k, \xi_{h^i})]} \Theta( f_{\alpha,L}(\xi_k, \xi_{h^i})).
\end{equation}
Indeed, the coupling strength between antigen and antibody is
calculated according to the rule introduced in Chapter
\ref{ch:structure} for antibody-antibody interaction since the forces underlying
the couplings are of the same nature. Therefore we can rewrite the
Hamiltonian $(2.7)$ as
\begin{equation}
H(\sigma,\mathbf{J},\mathbf{h}) = - \frac{1}{N} \sum_{k<j}^{N,N}
J_{kj} m_k m_j + \sum_{k=1}^N c \; h^i_k m_k,
\end{equation}
where $c$ can be possibly tuned to mimic variations in the antigen
concentration. This way, an arbitrary lymphocyte $k$ is subject to
two stimuli, one deriving from the presence of the antigen, and
the other from the presence of the remaining lymphocytes. This can
be formalized by saying that the field acting on the $k^{th}$ node
is
\begin{equation}
H_k = -  \frac{1}{N} \sum_{j=1}^N J_{kj} m_j + c \; h^i_k.
\end{equation}

In the absence of any antigen it is reasonable to consider all
lymphocytes in a quiescent state, i.e. $m_k = -1$ (under the
assumption of negligible noise) for any $k$; this provides the
initial state assumed to be stationary when no antigen is at work.
Hence, as the field $\mathbf{h^i}$ is switched on, we have
\begin{equation}
H_k =  \frac{1}{N} w_k - c \; h^i_k,
\end{equation}
where we used Eq.~(\ref{eq:connectivity}). Now, if $H_k$ occurs to
be negative, the state for the $k^{th}$ lymphocyte which minimizes
the energy is the firing one, namely $m_k=+1$. Hence, assuming
that all lymphocytes are quiescent, the condition for lymphocyte
$k$ to fire is
\begin{equation} \label{eq:cond_fire}
c \; h^i_k > \frac{w_k}{N}.
\end{equation}
This means that the minimal concentration necessary in order
to elicit an immune response by $k$ is directly proportional to
its degree $w_k$ and inversely proportional to its coupling
$h^i_k$. This also suggests that, in the presence of the antigen
$\xi_{h^i}$, the most reactive idiotype is not necessarily
$\xi_i$, but rather it may be a spurious one which exhibits the lowest ratio
$\frac{w_k}{h^i_k}$. Interestingly, the threshold mechanism
determined by Eq.~(\ref{eq:cond_fire}) consistently mimics the
low-dose tolerance phenomenon: The immune system attacks antigens
or, more generally proteins, if their concentrations  is larger
than a minimum value, which depends on the particular protein.

\begin{figure}[tb]
\begin{center}
\resizebox{0.55\columnwidth}{!}{\includegraphics{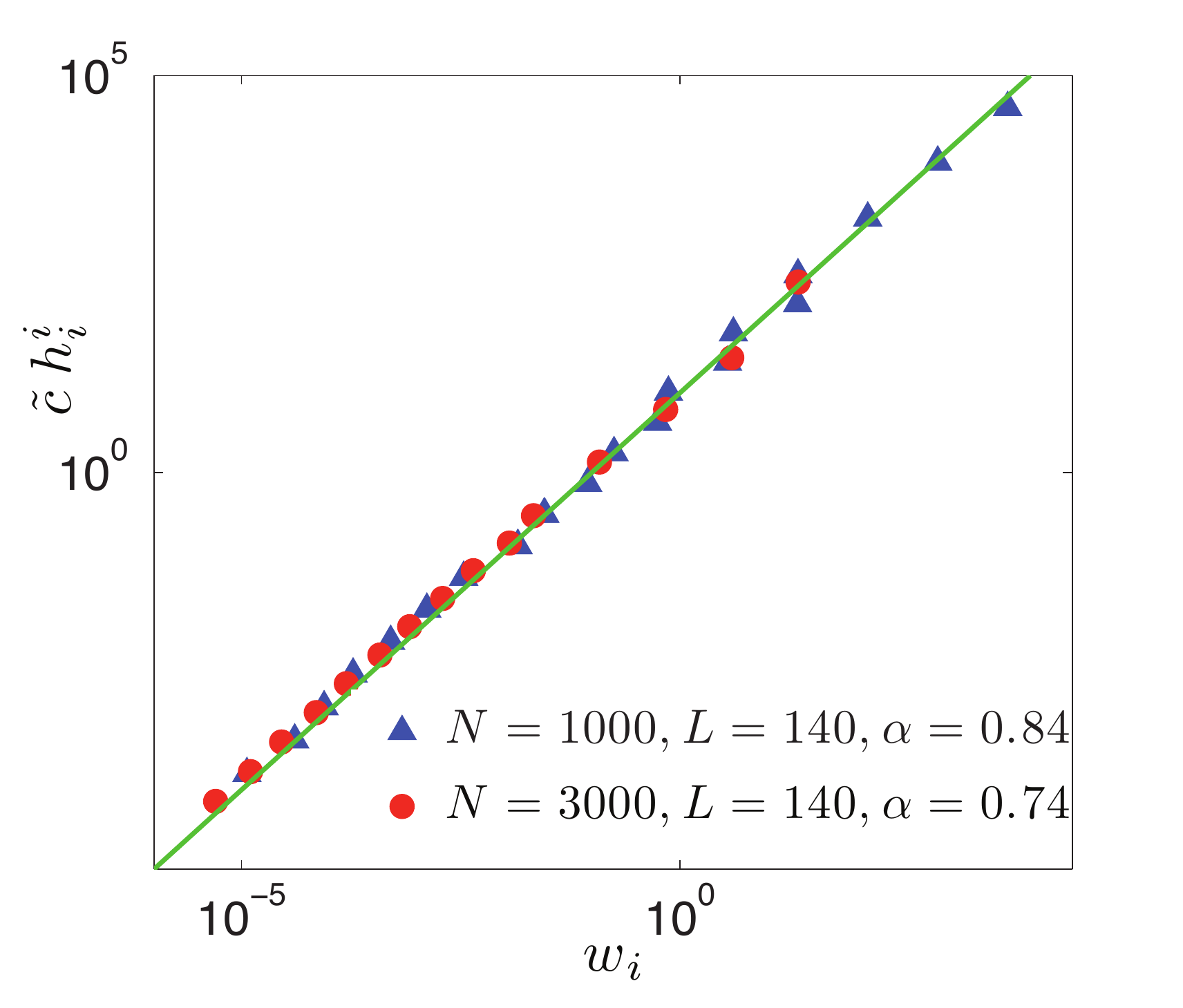}}
\caption{Threshold concentration $\tilde{c}$ of antigen $\xi_{h^i}$ as a function of the weighted connectivity $w_i$ of the first reactive family, in systems with parameters $N=1000$, $\alpha = 0.84$, $\langle k \rangle \approx 150$ (triangles) and $N=3000$, $\alpha = 0.74$, $\langle k \rangle \approx 40$ (circles), respectively; for both cases we used $M=100$ and $L=140$.  The level of noise is set as $\beta = 1.0$, far above the critical value.} \label{fig:threshold}
\end{center}
\end{figure}

Implicitly this mechanism suggests possible interpretation even of
the high dose tolerance: as what elicits a given lymphocyte is the
product of the weighted connectivity with another agent (antigen
or internal molecules) times its concentration (properly expressed
via a magnetization function), it can not distinguish among self
or antigen in an high dose.
\newline
Responding to high dose of antigen should, in principle, allow a
response even to the self, whose defence turns out to be a
primarily goal.
\newline
These analytical estimates have been checked by means of numerical simulations:
For a given system $(\alpha, L, \langle k \rangle )$ we run several experiments, each for a different applied field
$\xi_{h^i}$, where the concentration of the antigen is tuned from
$0$ up to the minimal value $\tilde{c}^i$ necessary to elicit an
immune response; data are reported in Fig. (\ref{fig:threshold}).
On the $x$-axis we set the weighted degree of the first reactive
lymphocyte and on the $y$-axis we set the minimal concentration $\tilde{c}$ of the
agent $\xi_{h^i}$ able to give rise to an immune response, multiplied by $h^i_i = \exp(\alpha L)$; a linear dependence between $w$ and $\tilde{c}$ is
evidenced by the fit, in agreement with Eq. (\ref{eq:cond_fire}).
We therefore recover the important result from Varela et al
\cite{a38,a39} that the reactivity of an antibody is closely
related to its degree, where, here, the degree is more
specifically meant as weighted degree.

The linear scale between $\tilde{c}$ and $w$ has some significant
consequences: The tolerated concentration of antigen directly
reflects the (weighted) inhomogeneity of the graph. Otherwise
stated, if the weighted degree spans a range, say, $\mathcal{O}(10^k)$, the tolerated concentrations relevant to all lymphocytes making up the system is expected to span an analogously wide range. Hence, recalling that
non-self-addressed Ig's belong to the left tail of the weighted-degree distribution, while self-addressed Ig's lay on the right tail, we have that the doses typically tolerated by the former are $k$ order of magnitude less that
those tolerated by the latter. Now, as shown in
Sec. \ref{sec:weight}, the weighted-degree distribution $P(w)$
exhibits a standard deviation scaling exponentially with $L$ or, analogously, algebraically with $N$, being $1/2$ a lower bound for the exponent. As a consequence, we expect that the region spanned by $w$ grows non-slower that $\sqrt{N}$; this means that for real systems the difference between doses tolerated by self and non-self is at least $\mathcal{O}(10^7)$.

We finally stress that the low-dose tolerance emerges as a genuine
collective effect directly related to the properties of the
idyotopic network and, in particular, on the distribution of the
weighted degree.


\section{Multiple responses and spurious states}

As explained in Sec.~\ref{sec:lowdose}, the introduction of a concentration $c$ of a given antigen described by the
external field $\mathbf{h^i}$ is able to increase the
magnetization (concentration) of the node (lymphocyte) $i$,
provided that $c$ is sufficiently high. In
general, if the concentration is large enough, several clones, different from $i$, may prompt a response: some of
them, say
$j_1,...,j_p$, (hereafter called spurious, once again in order to stress
similarities with neural networks \cite{a11}) respond because of a
non-null, though small, coupling with the external field, some
others, say
$j'_1,...,j'_p$, (hereafter Jerne states for consistency) respond because they display a strong interaction $J_{ij}$
with the formers or with the specific Ig $i$. Spurious
states can be very numerous according to the particular antigen
considered and to its concentration; under proper conditions the
response of spurious states can be even more intensive than the
specific response from $i$. In fact, the reactivity of a given
node $j$ is determined not only by the relevant antigenic stimulus
$h^i_j$, but also by its local environment, namely by the concentration of firing lymphocytes to which $j$ is connected.

While in a neural networks framework spurious states correspond to
"errors" during the retrieval (once a stimulus is presented) and
should be avoided, in immune network spurious states are
fundamental for an effective functioning of the whole machinery.
In fact, when an antigen is introduced in the body, all the set of
responders (proper lymphocyte and spurious states) do contribute
to attack the enemy and neutralize it. Moreover, the reaction of
spurious states can in turn have important consequences on the
generation of memory cells and on the effectiveness of the
secondary response. Accordingly, one can investigate whether it is
possible to figure out proper strategies which can limit or
increase the number the number of such spurious states. In
particular, how the connectivity of the structure and the coupling
patters $h^i_k$ influence this?

\begin{figure}
\resizebox{0.95\columnwidth}{!}{\includegraphics{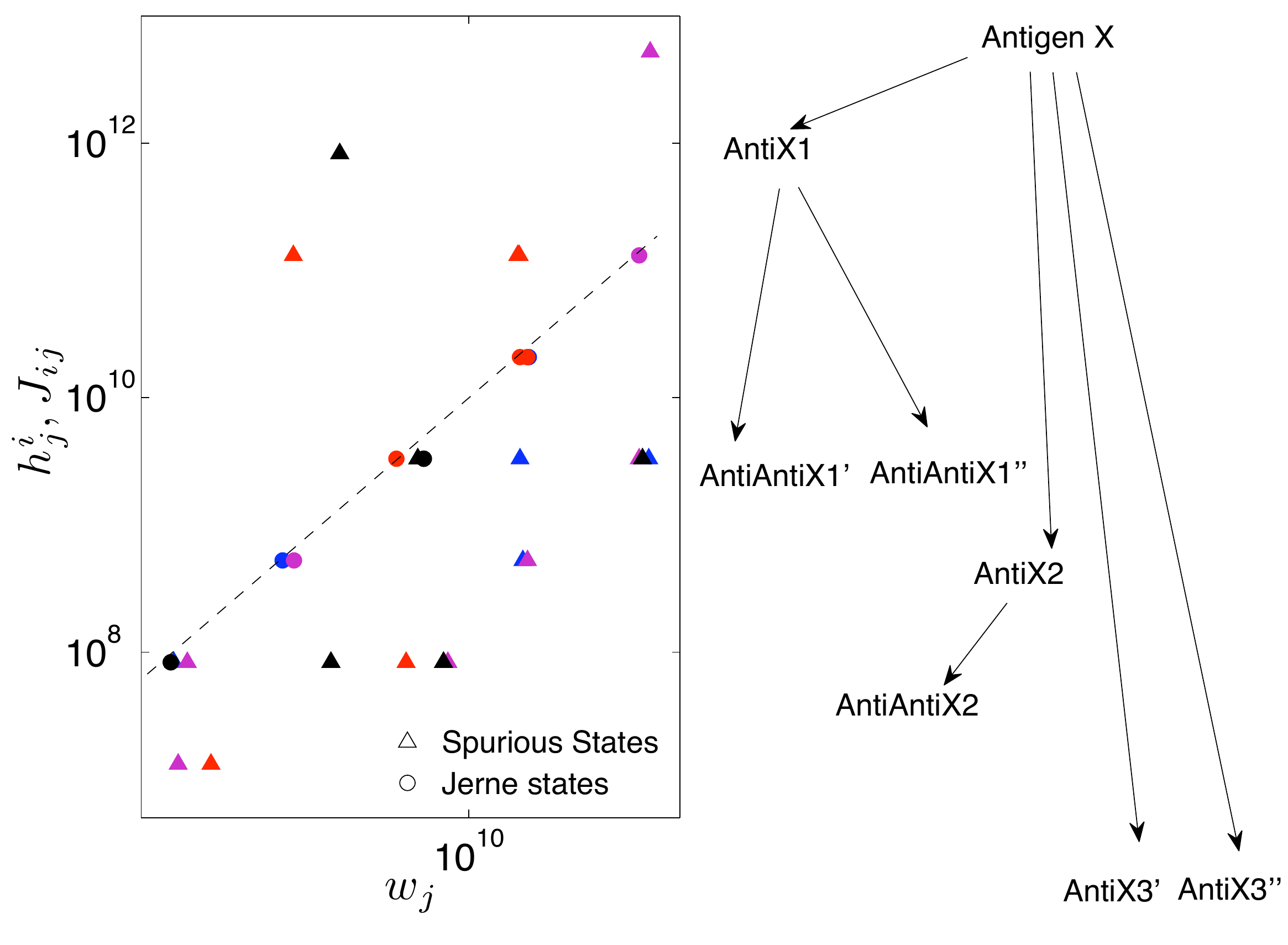}}
\caption{Analysis of spurious states for different realizations of a system made up of $N=1000$ families and $M=10$ clones per family; we assumed $\alpha = 0.84$ and $L=140$. Left panel: Field coupling $h^i_j$ (for reactive spurious states (triangles)) and interaction strength $J_{ij}$ (for reactive anti-antibodies (circles)) as a function of their weighted connectivity; each realization is depicted in a different color. Right panel: Schematic representation of connections between reactive antibodies.\label{fig:spuri}}
\end{figure}

Here, we just want to analyze the overall response of the system
outlining which kind of clone does react, that is, we distinguish between
spurious states and Jerne states, i.e. anti-Ig. Results for different
realizations of a system $(\alpha,L,\langle k \rangle)$ where the antigenic concentrations is set as $c \sim 10^2
\tilde{c}$ are shown in Fig. (\ref{fig:spuri}); different symbols
are used for spurious states (triangles) and for Jerne states (cirlces). For such concentrations the number of
reactive spurious states is approximately twice the number of
reactive Jerne states and a clear correlation between their
weighted connectivity and $J_{j'i,j}$ (Jerne state) can also be evidenced. Interestingly, Jerne
states require a larger stimulus to react and this is due to the fact that
the reaction is not directed, but rather mediated by the specific Ig $i$ or by a spurios state $j$.

Furthermore it is also important to stress that apparently,
without the introduction of spurious states, the amount of
antibodies is greater than the amount of lymphocytes and this would be in conflict with the first postulate of immunology: consistency is
obtained thanks to the lack of a perfect match among antibodies
(or antibody and antigen), which allows multiple attachments
ultimately accounting for a large over-counting of different
responses.

\section{Dynamical Memory}

As well known, the immune system is able to develop memory
effects; for this to happen the mutual interaction among
lymphocytes is crucial. The activation of a lymphocyte $i$ must
therefore be followed by the activation of the relevant
anti-antibody $\bar{i}$, which, in turn, may elicit the
anti-anti-anti-body $\bar{\bar{i}}$ and so on in a cascade
fashion. It is just the modulation and mutual influence among such
interacting antibodies to keep the concentration of antibodies
themselves at appropriate levels, which provides memory storage
within the system. In this section we want to analyze under which
conditions, if any, a firing state of lymphocyte $i$, i.e.
$m_i=1$, can determine a non-null concentration for the
anti-anti-body $\bar{i}$ to react.

Let us consider a system characterized by parameters $\alpha$, $L$
and $\langle k \rangle$, in such a way that $N$ is determined by
Eq. \ref{eq:scaling}. Assuming that $m_i=1$ and $m_k=-1$ for any
$k \neq i$, we have that , all in all, node $\bar{i}$ subject to a
field $H_{\bar{i}}$ given by the presence of the others $N-1$
families
\begin{equation}
H_{\bar{i}} = - \frac{1}{N} \left( \sum_{j \neq \bar{i}} J_{ij}
m_j + J_{i \bar{i}} m_i \right).
\end{equation}
which can be rewritten as
\begin{equation}
H_{\bar{i}} = - \frac{1}{N} \left[ - (w_{\bar{i}}-J_{i \bar{i}}) +
J_{i \bar{i}} \right],
\end{equation}
where we used $\sum_i J_{ij} = w_i$. We therefore derive that
$\bar{i}$ is also firing if
\begin{equation} \label{eq:conditio}
w_{\bar{i}} < 2 J_{i \bar{i}}.
\end{equation}
As shown in Sec.~\ref{sec:recognition}, anti-antibodies
corresponding to non-self addressed Ig's typically belong to the
so called mirror-block of the affinity matrix and they display and
average connectivity $w_{\bar{i}} \approx \langle J
\rangle_{\alpha,L} (N-1)$. Moreover, the affinity between $i$ and
$\bar{i}$ can be estimated as $J_{i \bar{i}} \approx \langle J
\rangle_{\alpha,L} + 2 \sigma^J_{\alpha,L}$, since the coupling
between Ig and Anti-Ig lays on the right tail of the coupling
distribution. Therefore, recalling $\langle J \rangle_{\alpha,L}
=1$, we can rewrite Eq.~\ref{eq:conditio} as
\begin{equation} \label{eq:conditio2}
N -1 < 2 (1 + 2 \sigma^J_{\alpha,L}).
\end{equation}
Now, we can use Eqs.~\ref{eq:scaling} and \ref{eq:J2_ex} to write the previous
expression as a function of $L$, $\alpha$ and $\langle k \rangle$:
\begin{equation} \nonumber
 \frac{2 \langle k \rangle}{\mathrm{Erf}\left( \sqrt{\frac{L}{2}} \right) - \mathrm{Erf}\left( \tilde{\alpha} \sqrt{\frac{L}{2}} \right) } < 3 + 4 \sqrt{2 e^{\frac{L}{4}(\alpha+1)^2 } }  \hspace{3.2cm}
 \end{equation}
\begin{equation} \label{eq:phasediag}
 \hspace{2cm} \times \frac{ \sqrt { \mathrm{Erf} \left( \frac{\alpha(3+\alpha)}{1 + \alpha} \sqrt{\frac{L}{2}} \right) - \mathrm{Erf} \left( \alpha \sqrt{\frac{L}{2}} \right) } }{  \mathrm{Erf}
\left(\frac{\alpha^2 + 4 \alpha -1}{2(1+ \alpha)}
\sqrt{\frac{L}{2}} \right) +   \mathrm{Erf}
\left( \frac{1 - \alpha}{2} \sqrt{\frac{L}{2}} \right)  } .
\end{equation}

\begin{figure}[tb]
\begin{center}
\resizebox{0.60\columnwidth}{!}{\includegraphics{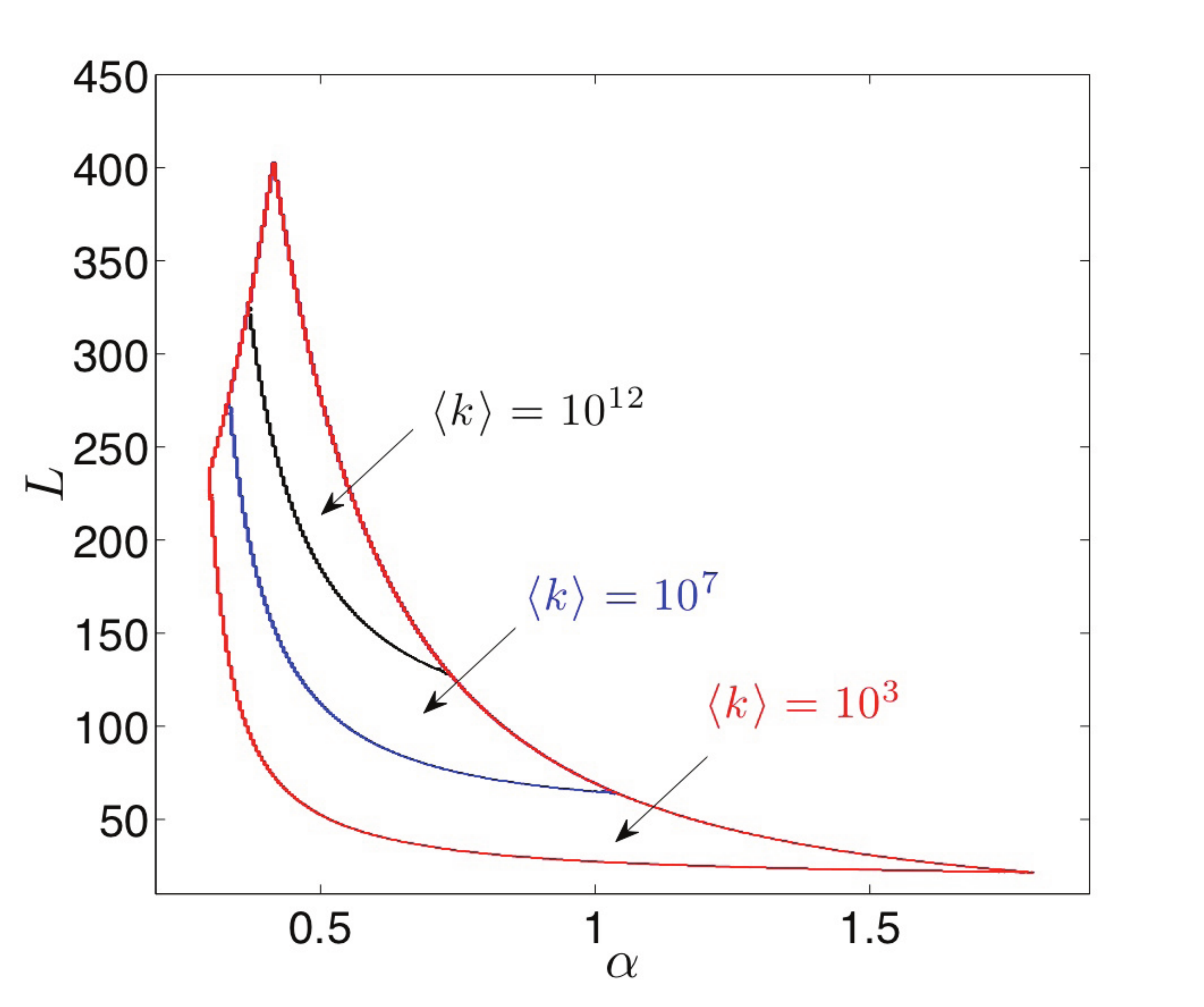}}
\caption{Numerical solution of Eq.~(\ref{eq:phasediag}): different values of $\langle k \rangle$ correspond to different closed curves, whose internal region provides the values of $\alpha$ and $L$ which satisfy the inequality of Eq.~(\ref{eq:phasediag}), namely the region of retrieval.} \label{fig:phasediag}
\end{center}
\end{figure}

A better insight in the previous expression can be achieved from Fig.~\ref{fig:phasediag} which shows its numerical solution for different values of $\langle k \rangle$ (each shown in a different color):
for a given average degree $\langle k \rangle$, the region of the plan $(\alpha,L)$ contained within the pertaining satisfies Eq.~(\ref{eq:phasediag}).
For instance, let us assume $\langle k \rangle = 10^{12}$ and
$L=140$, then $\alpha$ must be not larger than approximately $0.75$ if
we want that a response from the anti-anti-Ig follows the reaction
of a specific Ig.
By analyzing Fig.~\ref{fig:phasediag}, wenotice that the less diluted the network the smaller the region of ``retrieval''; Indeed, if we fix a given point on the $(\alpha,L)$, increasing $\langle k \rangle$ implies a larger $N$ and this contrasts with the satisfability of  Eq.~(\ref{eq:phasediag}), on the other hand this result is rather intuitive as, when the coordination is large, the anti-antibody is less reactive with respect to the stimulus provided that the related antibody. As for the region corresponding to large $L$ and relative large $\alpha$, this never overlaps with the retrieval region, in fact for those values $\sigma_{\alpha,L}^J$ is relatively small.

Finally, we stress that when $\langle k \rangle$ is close to real values, i.e. $\mathcal{O}(10^{12})$ the retrieval region is centered on $L \sim 10^2$ and $\alpha$ between $0.5$ and $1$, this is consistent with the conclusions drawn from the analysis of the network dilution (Sec.~\ref{sec:dilution}) and from the analysis of sel/non-self recognition (Sec.~\ref{sec:recognition}).

\section{Generation of memory cells}

The last step we perform is to zoom inside the immune system and
consider the behavior of single clones when stimulated with fields
so to secrete  immunoglobulins  in high concentration; to this
task, as the antigen concentration varies during the infection (at
first increasing than -hopefully- decreasing) we have to deal with
time dependent fields pushing the system away from the
equilibrium.
However for the sake of clearness at first we work out the
equilibrium behavior of only two clones in interaction with each other and with their corresponding fields, so to have
equilibrium even at this zoomed level, then, in the other
sections, we will study their dynamical features.

Another interesting property of the immune system is the long-term
memory of the past infections: When an antigen has been
successfully rejected, part of the stimulated lymphocytes
(corresponding to clones which have undergone expansion) do not
recover the rest but they become "memory-cells", namely they
remain activated, such that if the same antigen is re-introduced
in the body an immediate, and highly specific response, usually
improved with respect to the first time, can raise immediately.

The amount of memory cells achieved after an infection is not a
constant and it is known to depend on the time of the infection
 and on the kind of the infection \cite{a11,a101}.

We want to show that our model exhibits a phenomenon called
"hysteresis" \cite{a61,a62} and this may account for the
generation of the memory cells, furthermore it can model naturally
both the improvement of the quality of the antibody production in
the second response and the time-dependence in the ratio of the
obtained memory-cells.

The hysteresis is a dynamical feature (disappearing in the
quasi-static limit \cite{a60}) which essentially arises due to
conflicting timescales inter-playing.

In ferromagnetic materials hysteresis is a well known phenomenon,
both theoretically and experimentally: When an oscillating
magnetic field is applied to a ferromagnet, the thermodynamic
response of the system, namely its magnetization, will also
oscillate and will lag behind the applied field due to the
relaxational delay. Therefore, the two time-scales inter-playing
are given by the frequency of the external field $\omega$ and by
the thermalization of the system itself. When relaxation is slower
than field oscillations we have a delay in the dynamic response of
the system which gives rise to a non-vanishing area of the
magnetization-field loop; this phenomenon is called \emph{dynamic
hysteresis}.

In an immune system the two dynamical phenomena and related
time-scales involved are: the raise of the immune response (which
may depend by the particular clone interested) and the antigenic
growth (which strongly depends by the given antigen). When the two
subjects are made interacting delays in the dynamic response gives
rise to several interesting phenomena: at first when the
concentration of the antigen is bring back to zero at the end of
the fight, there can be not zero -remanent- magnetization (lacking
of the in-phase behavior appears), further, if we deal with
periodic perturbation, when the time period of such antigenic
oscillations becomes much less than the typical relaxation time of
the clone, a dynamical phase transition toward a chronic response
may appear.
Ultimately this approach may bridge the Barkhausen effect
\cite{a60} to the "Jerne avalanches" of antibody and
anti-antibody, whose distribution and sizes are interesting
information.

Before results are outlined, we stress that we applied sinusoidal
fields of the form $h(t)=h_0\sin(\omega t)$, where $h_0$ takes
into account the interaction thought the lymphocyte network, while
$\omega$ rules the timescale of the antigen.

\section{Two clones dynamics and maturation of secondary response}

\begin{figure}[tb]
\resizebox{0.45\columnwidth}{!}{\includegraphics{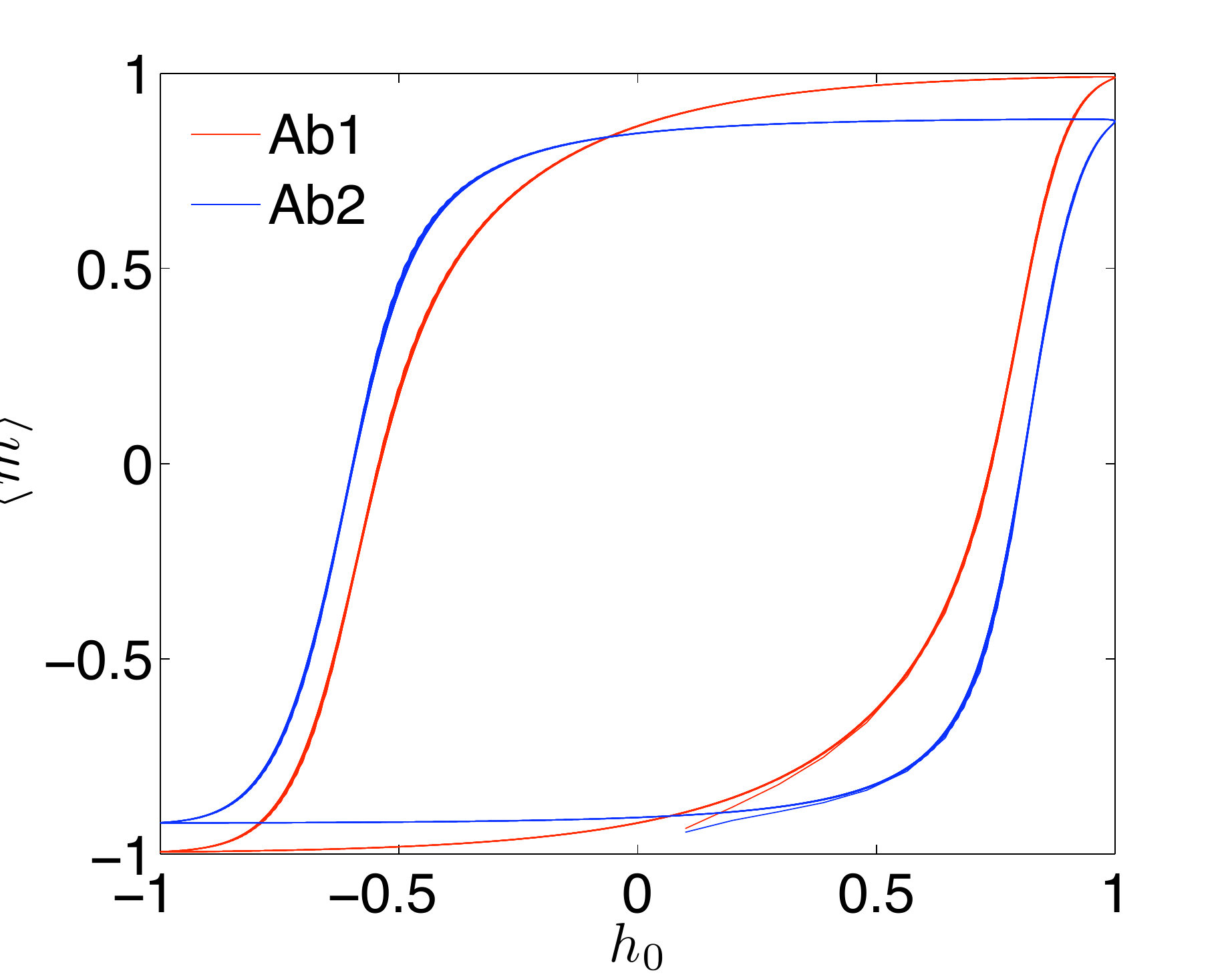}}
\resizebox{0.45\columnwidth}{!}{\includegraphics{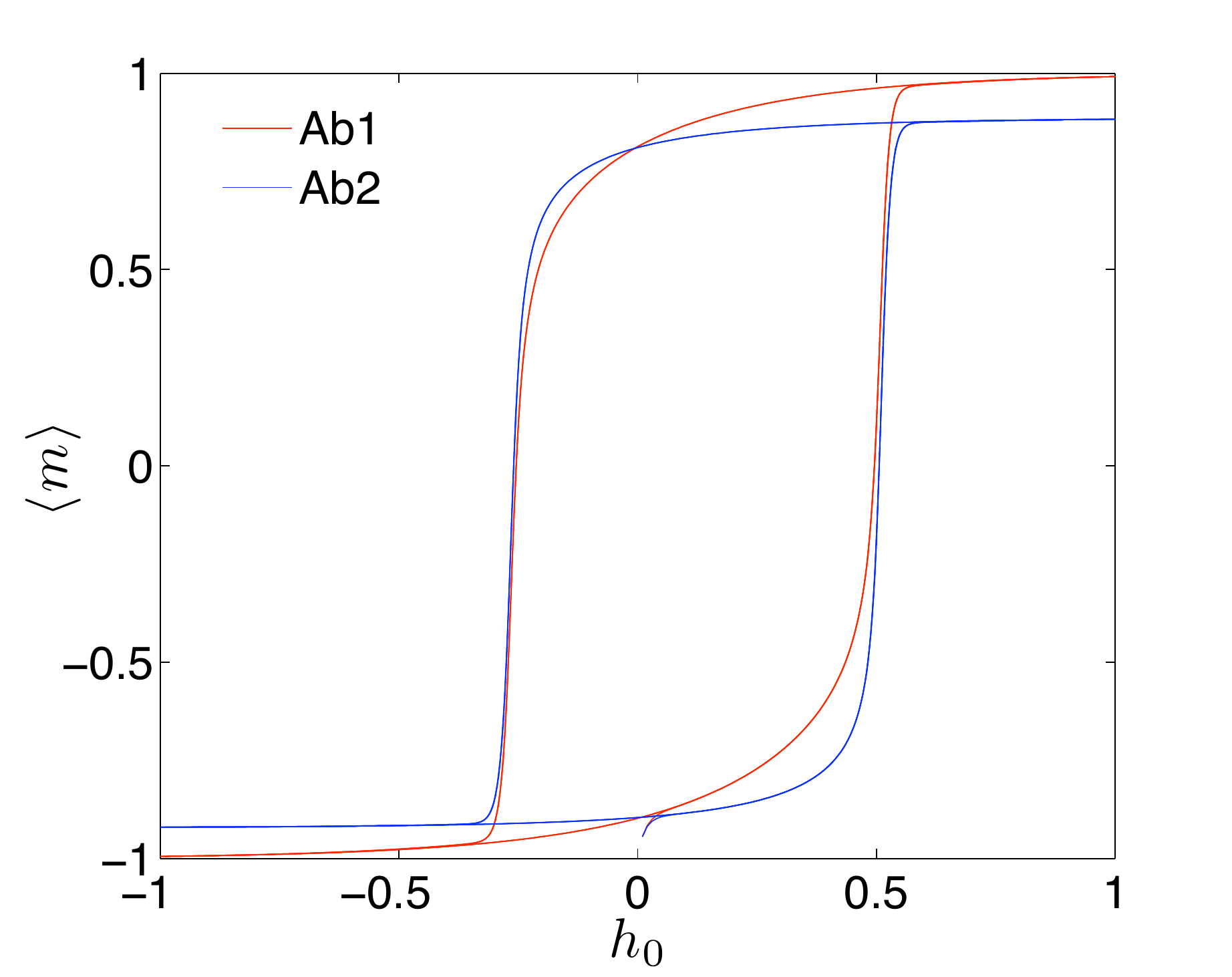}}
\caption{Examples of hysteresis cycles for a two-body system: on the $y$-axis we show the concetration of firing lymphocytes and on the $x$-axis the amplitude of the external field. Two different frequences are considered: $\omega = 0.1$ (left panel) and $\omega=0.01$ (right panel). The level of noise is below the critical value.}
\label{fig:isteresi}
\end{figure}

\begin{figure}[tb]
\begin{center}
\resizebox{0.5\columnwidth}{!}{\includegraphics{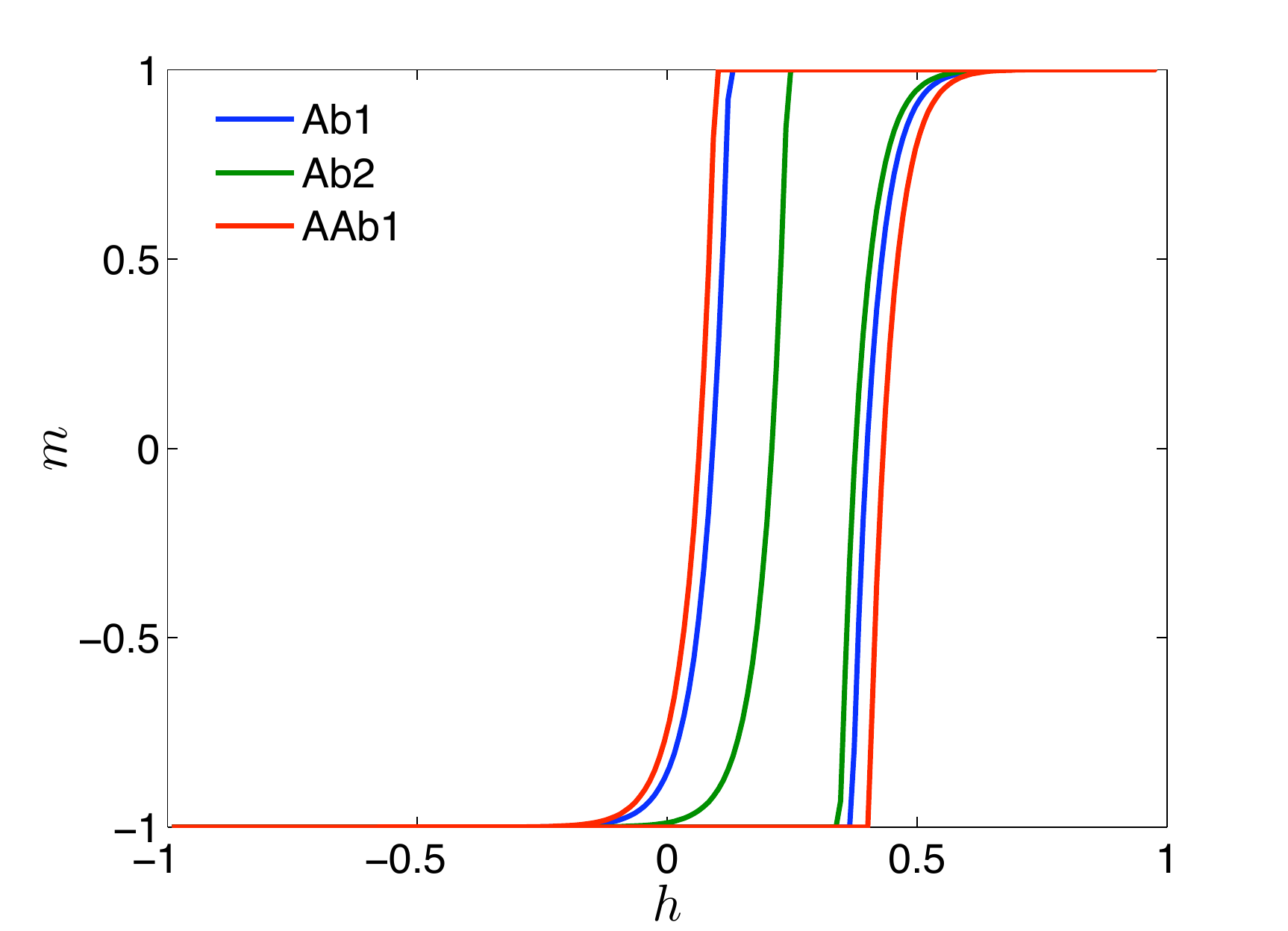}}
\caption{Hysteresis displayed by the stimulated best matching
clone Ab1, a spurious clone Ab2 and the Jerne counterpart of Ab1,
namely AAb1. From the picture the maturation of best performing
response can be understood as follows: at the first infection with
the antigen both the best matching clone (Ab1) and a spurious
state (AAb1) are elicited. However Ab1 experiences an higher field
such that its saturation in the hysteresis curve is greater with
respect to Ab2 and so is its remanent magnetization. At the second
infection of the same antigen the quiescent state is no longer the
old one as now different concentrations of the two clones are
stored and in particular Ab1 shows higher values than AAb1 so its
response starts immediately higher and overall the immune response
is stronger, due to both the remanent magnetization, and sharper,
due to the mismatch among their relative concentrations.}
\label{fig:isteresi}
\end{center}
\end{figure}

For simplicity, let us start considering only two clones ($N=2$)
in interaction, as the generalization to slightly more involved
situations (i.e. four clones) is straightforward. The macroscopic
states of the two clones are
 $m_1 = M^{-1}\sum_{\alpha=1}^{M}
\sigma_1^{\alpha}$ and $m_2=M^{-1}\sum_{\alpha=1}^{M}
\sigma_2^{\alpha}$.
\newline
Given these two clones, $i$ and $j$, their mutual interaction
parameter $J_{ij}$ depends on the subset they belong to, as
specified by the symmetric matrix
\begin{displaymath}
         \begin{array}{ll}
                \\
                P \left\{ \begin{array}{ll}
                \\
                \\
                                  \end{array}  \right.
                \\
                P \left\{ \begin{array}{ll}
            \\
                \\
                                  \end{array}  \right.
         \end{array}
          \!\!\!\!\!\!\!\!
         \begin{array}{ll}
        \quad
                 \overbrace{ \qquad \qquad }^{\textrm{$P$}} \ \ \
                 \overbrace{\qquad \qquad}^{\textrm{$P$}}
                  \\
                 \left(\begin{array}{cc|cc}
                               \mathbf{ J_{11}=0}  &  & \mathbf{ J_{12}=J}
                                \\
                                \\
                                \\
                                \mathbf{ J_{12}=J} &  & \mathbf{ J_{22}=0}
                                \\
                      \end{array}\right)
               \end{array}
\end{displaymath}
where each matrix block has constant elements:
$J_{11}=J_{22}=0$, as lymphocytes carrying the same idiotype do
not interact, while  $J_{12}=J_{21}>0$ controls the interaction
between lymphocytes of different clones.

Analogously, the field $h_i$ takes two values $h_1$ and $h_2$,
depending on the type of $i$, as described by the  following
vector:
\begin{displaymath}
         \begin{array}{ll}
               M \left\{ \begin{array}{ll}
                                      \\
                                      \\
                                   \end{array}  \right.
                                        \\
                M \left\{ \begin{array}{ll}
                                          \\
                                         \\
                 \end{array}  \right.
           \!\!\!\!\!\!
    \end{array}
    \!\!\!\!\!\!
    \left(\begin{array}{cc|cc}
                \mathbf{h_{1}}
            \\
            \\
            \mathbf{h_{2}}
            \\
            \\
        \end{array}\right).
\end{displaymath}
This model, two paramagnets ferromagnetically interacting, has two
coupled self-consistence relations \cite{a43,a100} which solves
the $t \to \infty$ limit of the stochastic dynamics, worked out as
\begin{eqnarray}\label{PL1}
\frac{d m_1(t)}{dt} = - m_1(t) +  \tanh\Big(\beta (J_{12} m_2(t)  + h_1(t) )\Big), \\
\frac{d m_2(t)}{dt} = - m_2(t) +  \tanh\Big(\beta (J_{12} m_1(t)
+h_2(t))\Big).\label{PL2}
\end{eqnarray}
Despite its simplicity the phase diagram of these two interacting
clones is already rich of both continuous and discontinuous
transitions \cite{a43,a100} and is found to have not trivial
stable concentrations of the clones accounting for the optimality
of the free energy density.

\begin{figure}[tb]
\resizebox{7.5cm}{6.5cm}{\includegraphics{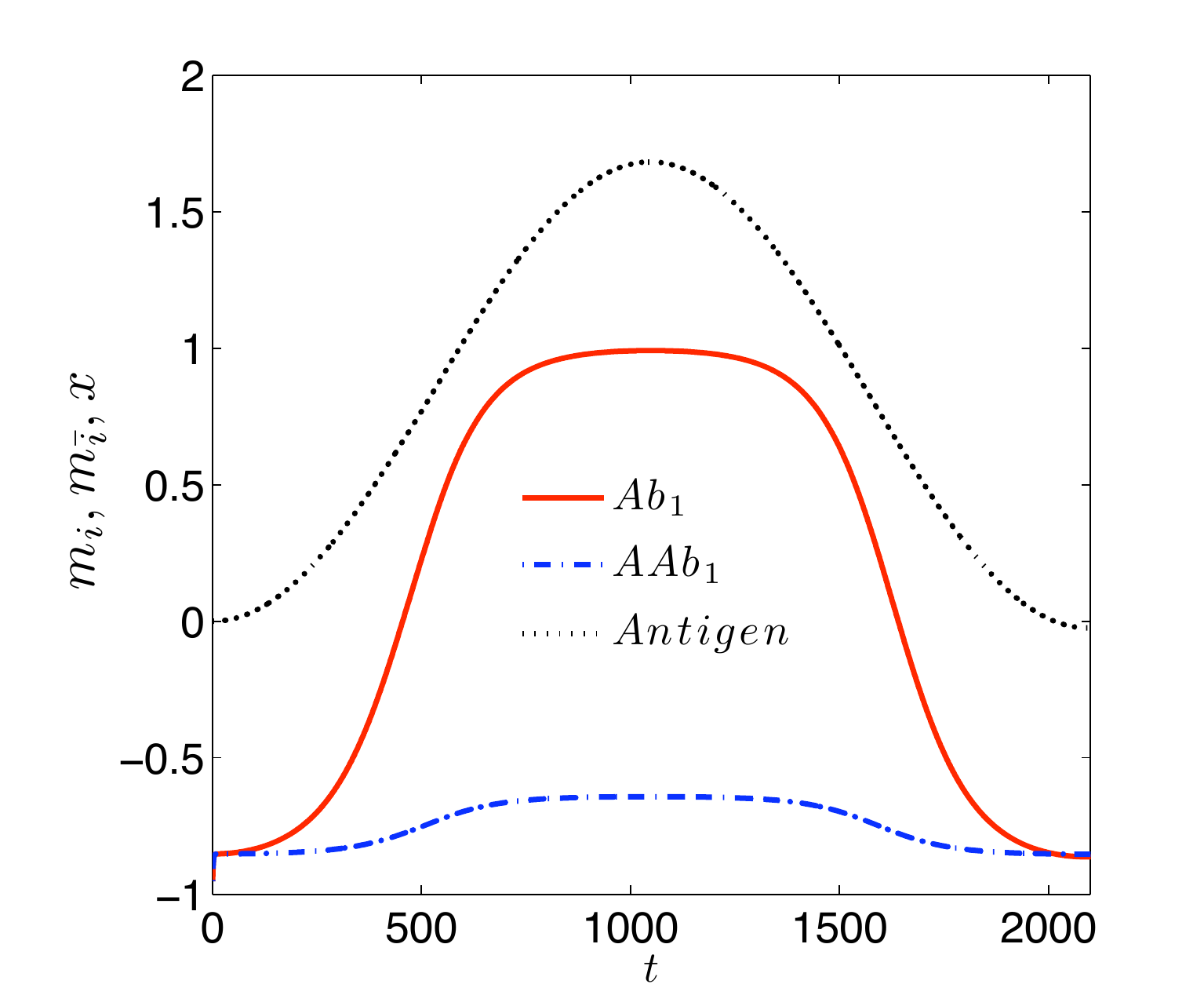}}
\resizebox{7.5cm}{6.5cm}{\includegraphics{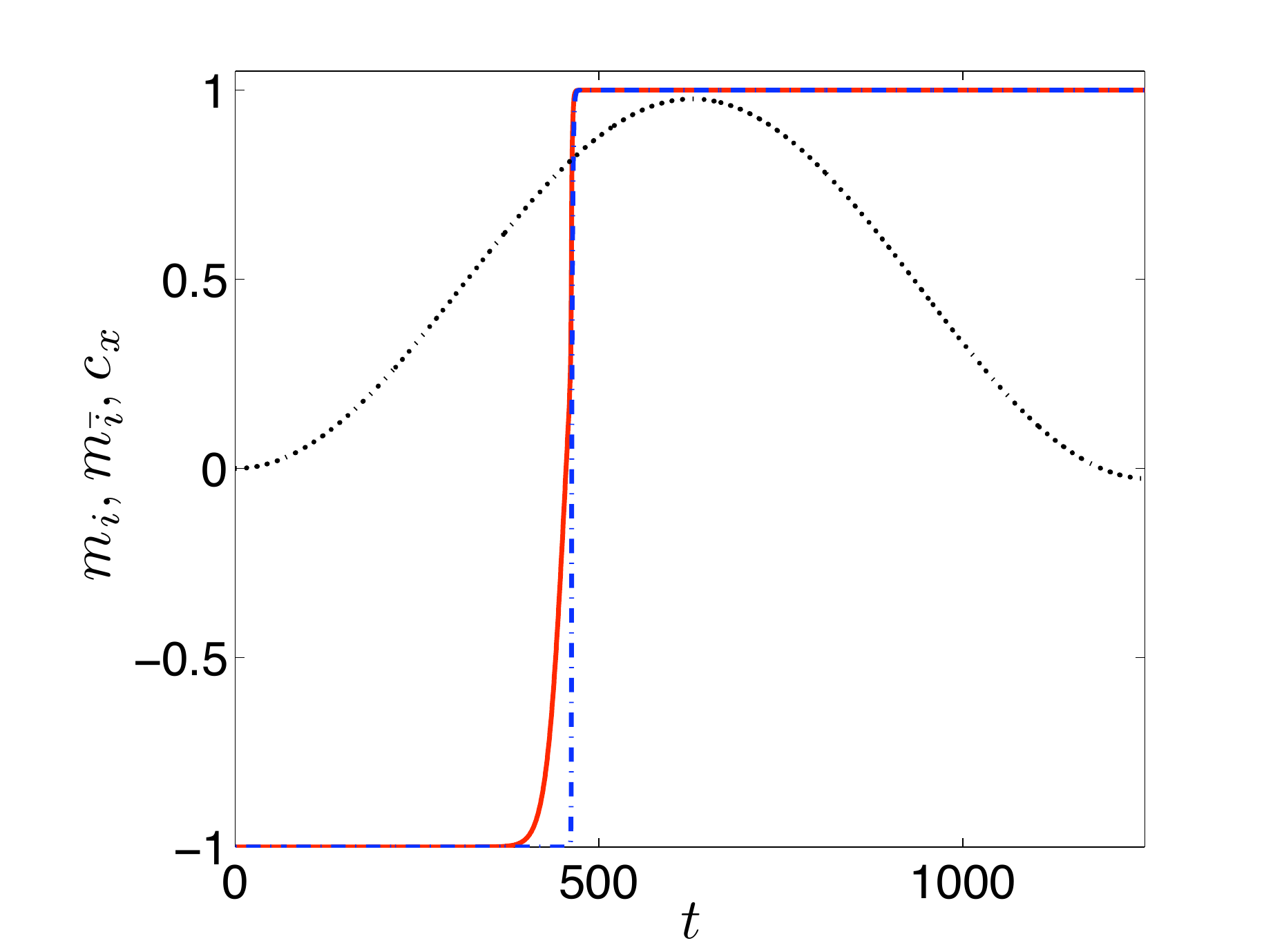}}
\caption{Examples of bell-shaped response for a $2$-clones system} \label{fig:bell2}
\end{figure}

To accomplish our task we start applying  a field $h$ to the
immune system at rest and collect the responding clones.

For these clones we generalize the Langevin equation of the
two-body model (eq.s \ref{PL1},\ref{PL2}) so to obtain a system of
coupled stochastic equations (one for each order parameter of a
stimulated clone) that we integrate via the step adaptive
Runge-Kutta algorithm \cite{a107}.
\newline
Roughly speaking, if we define -as a measure of the out-of-phase
response- the area of the hysteresis loop in the $h_i,m_i$ plane
as
$$
A(\beta,h_0,\omega) = \oint m dh
$$
we found that this area is increasing with $\omega$ at low
frequencies (because increasing the frequency increases the
delay), then reaches a maximum and than start decreasing (due to
the $2\pi/\omega$ periodicity); when looking at this area versus
the noise level it is seen to increase when $\beta$ increases
(that means that the affinity matrix can be more felt by the
system and consequently its delay increases because of the storing
of information inside the relative coupling concentrations). In
figure (\ref{fig:isteresi}) we show a typical behavior (for
$\beta<\beta_c$) of the first two best fitting clones elicited by
an external antigen: A proposal for the understanding of the
improvement of both the quantity and the quality of the secondary
response can be obtained from the picture.

The secondary immune response is stronger because the best fitting
lymphocyte (Ab1) at the second infection do not start off from the
minimal allowed value (as the first time) but from the value of
concentration related to the remanent magnetization.
It is also more specific than the first response. This can be
understood by the following argument: At the beginning all the
clones start off from the quiescent values. However, the
lymphocyte with the best fitting antibody experiences an eliciting
field stronger than the others (in particular than AAb1) and so it
reaches saturation in the hysteresis for a longer time. This
allows a greater remanent magnetization with respect to the
spurious state which expanded as well. As a consequence, if the
stimulus is presented once again, the immune system is not simply
translated from quiescence level of concentrations to remanent
magnetization levels, but different values of the latter, among
Ab1 and AAb1, account for an improved response.

\section{Bell shaped response}

In this section we want to show that, within our model, the so
called "bell-shaped response" \cite{a11} is recovered as the
typical immune response.

As we are interested in basic features of the immune system, we
look at the two most common responses to only the positive half of
a sinusoidal stimulus: the two-clone circuit and the four-clone
circuit, which naturally extend the Eqs. (\ref{PL1}, \ref{PL2}),
the only difference being the coupling field which by now acts
only on $m_1$ (whose concentration we want to measure).

\begin{figure}[tb]
\resizebox{7.5cm}{6.5cm}{\includegraphics{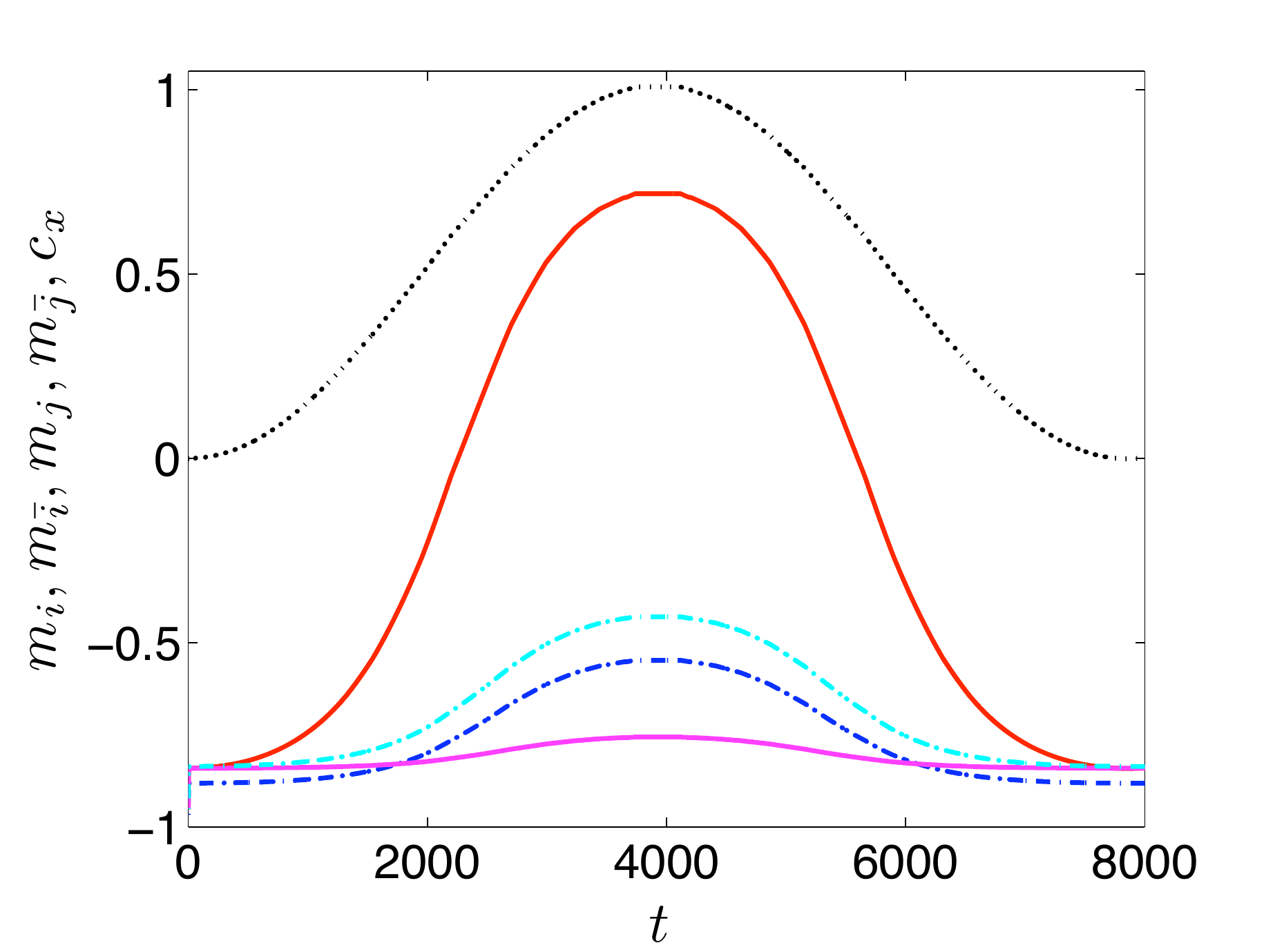}}
\resizebox{7.5cm}{6.5cm}{\includegraphics{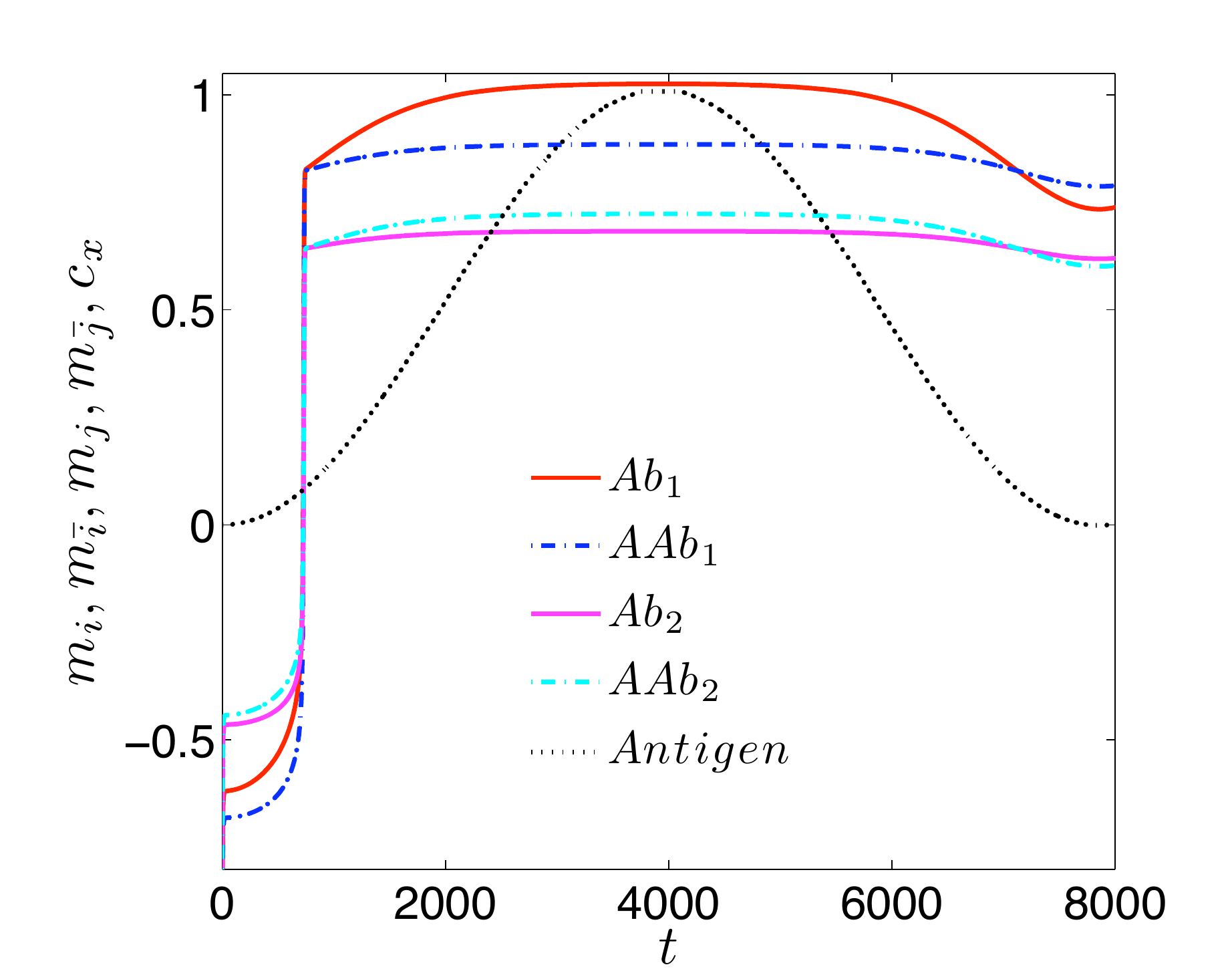}}
\caption{Examples of bell-shaped response for a $4$-clones system} \label{fig:bell4}
\end{figure}

Before presenting our results it is worth spending few lines to
introduce more clearly our approach: usually when dealing with
these dynamical features of the immune system, the most intuitive
way is to work out time-differential equations for the evolution
of the antigenic and antibody concentrations (which is indeed what
is almost always done \cite{a79,a108,a109}. This account for
$m(t)$ and $h(t)$, then one, carefully looking at the monotonic
regions of the two behaviors, can parameterize and get $m(h)$.
Within our approach we do not solve this kind of problem, instead
we directly give an expression for the antigenic load and then we
obtain $m(h)$ as a result. This has two advantages: this allows to
deal with Fourier analysis, furthermore this skips all the
troubles about the details of the interactions which may strongly
depend on the particular antigen \cite{a11}. Of course we pay the
price of testing these responses with ideal mono-frequency viruses
which are surely an oversimplification: however a more complex
behavior can be obtained considering the antigen as a sum of
several perturbing harmonics, which is allowed due to the
linearity of the Hamiltonian $H_1$ with respect to the fields.

Typical results of bell-shaped found at different field frequency and magnitude are presented in the panels of
Fig.~\ref{fig:bell2} and of Fig.~\ref{fig:bell4} .


\chapter{Conclusions and Perspectives}

\section{Summary}

In this paper we pioneered an alternative way to theoretical
immunology by plugging it into a concrete disordered statistical
mechanics framework: our work is not meant as an exhaustive picture
of the (adaptive response of the) immune system, but rather as a
starting point in modeling its universal features by means of
this technique.

We stress that, in our model, once the amount of available
epitopes and their distributions is given (namely the amount of
$\xi$'s together with their distribution), everything can be
worked out. This way, we recover, qualitatively and partially
quantitatively,  all the basic known features of the immune
system and we obtain a good agreement with experimental data,
starting with a reasonable amount of working lymphocytes and
antibody concentrations.

In particular, in the complex system framework we developed, the
immune network naturally and autonomously achieves/recovers the
following properties:

\begin{itemize}

\item the Burnet clonal expansion theory \cite{a6}\cite{a48}  appears as the
standard one-body response of the system described by the
Hamiltonian $(2.5)$ (remembering the bridge among magnetization
and concentration encoded into eq. $(2.4)$).


\item the multi-attachment among antibodies is a natural property
of the system and gives raise to the Jerne network (cfr.
eq.(\ref{eq:J}))

\item the Jerne antibody network, which is obtained as a random
graph, encodes dynamically the memory of the encountered antigens
(see fig.$(3.4)$ and fig. $(4.4)$).

\item the Varela-Counthino self/non-self distinction appears as an emerging
property of such a network (see fig. $(3.6)$ and fig. $(4.1)$).

\item the Low  Dose Tolerance is the inertia of the network when
subjected to respond to external fields (see fig. $(4.2)$).

\item the existence of several antibodies acting against a given
antigen,  play the role of the dynamical spurious states of the
neural network static counterpart (see fig. $(4.3)$).

\item the High Dose Tolerance appears as a  mechanism avoiding
the breaking of self-recognition (see sec. $(4.2)$).

\item the genesis of memory cells (accounting for the transition
IgG -> IgM in antibody secretion) is played by the hysteresis in
the network (see fig. $(4.5)$).

\item the bell-shaped function as immune response is an emergent behavior and
not a postulate (see fig. $(4.7)$ and fig. $(4.8)$).

\item the secondary response is stronger and better than the
first (see fig. $(4.6)$).

\item increasing the noise, both the the quality
and the quantity  of the available retrievals decrease.

\end{itemize}

All these different aspects of the immune system appear as
features of a whole unified quantitative and very simple theory.
Furthermore, over  these agreements, matching with the
experimental data is available: namely the average connectivity of
the network is in agreement with the experiments as well as the
reciprocal affinities of the cascade of complementary antibodies.

With purely physical eyes our model describes the equilibrium of
the immune system as a (thermodynamically \cite{a50}) symmetry
broken random bond diluted ferromagnet. However, its non
equilibrium states map the latter into a random field random bond
diluted model, conferring to the system a glassy flavor.

\section{Outlooks}

Among the several outlooks surely the out of equilibrium
thermodynamics is to be obtained as the model is shown to display
a very rich ensemble of timescales and aging is expected. Another
important point is its learning, that so far is left
uninvestigated, which merges the approach of neural networks
\cite{a24} and the dynamical graph theory with information theory.
The extension of the concept of Hopfield statical memories into a
dynamical counterpart should be deepened as well as the Gardner
saturation bound \cite{a14}, which may play a key role in the
breaking of defense in the body. The transition from a "simple"
system to a "spin glass" due to the increase of pasted random
fields also needs a deep analysis as it is concerned with the
genesis of autoimmune responses. The interplay among B cells and T
helper cells should also be taken into account as T helper play
the role of a spin glass self-regulation adding a considerable
amount of complex self-regulations. At the end, as our results are
qualitatively quite robust and the framework very stable under the
change in the epitope distributions, other, mathematically
challenging (i.e. due to correlations), choices for these
distributions are surely biological plausible and should be
investigated. We plan to report soon on several of the outlines
directions of research.

\section*{Acknowledgment}
The author are grateful to O. Barra, R. Burioni, M. Casartelli, P. Contucci, A.C.C. Coolen, S. Franz, F. Guerra,  G. Ruocco and N. Shayeghi for useful discussions.


\begin{thebibliography}{}


\bibitem{a3} F. Celada, {\em Search of T-cell help for the
internal image of the antigen}, Theor. Imm. \textbf{2}, A.S.
Perelson Ed. Addison-Wiley Pubbl. (1988). 

\bibitem{a4} V.G. Nesterenko, {\em Symmetry and asimmetry in the immune
network}, Theor. Imm. \textbf{2}, A.S.
Perelson Ed. Addison-Wiley Pubbl. (1988). 

\bibitem{a5} L.A. Segel, A.S. Perelson, {\em Computations in shape space: A new approach to
immune network theory}, Theor. Imm. \textbf{2}, A.S. Perelson Ed.
Addison-Wiley Pubbl. (1988).

\bibitem{a6} F.M. Burnet, {\em The clonal selection theory of acquired
immunity}, Vanderbilt Univ. Press. Nashville, (1959).

\bibitem{a7} F.M. Burnet, F. Fenner, {\em The production of
antibodies}, MacMillan Ed.s, Melbourne (1949).

\bibitem{a8} P. Elrich, {\em Studies in Immunity}, Wiley, New
York, (1910).

\bibitem{a9} W.J. Dreyer, J.C. Bennett, {\em Molecular bases of antibody formation: a
paradox}, Nobel Prize Lecture, (1965).

\bibitem{a10} G.W. Hoffmann, {\em A theory of regulation and Self-NonSelf discrimination in an immune
network}, Eur. J. of Imm. \textbf{5}, 638-643, (1975). 

\bibitem{a11} A. K. Abbas, A. H. Lichtman, J. S. Pober, {\em Cellular and molecular
immunology}, Elsevier Ed.s, (2007).


\bibitem{a13} E. Agliari, A. Barra, F. Camboni, {\em Criticality in diluted
ferromagnets}, J. Stat. Mech., (2008).

\bibitem{a14} D.J. Amit, {\em Modeling brain function: The world of attractor neural
network} \ Cambridge Univerisity Press, (1992)

\bibitem{a15} D.J. Amit, H. Gutfreund, H. Sompolinsky,  {\em Storing infinite numbers of patterns in a spin glass model of neural
networks}, Phys. Rev. Lett. \textbf{55}, (1985).


\bibitem{a17} R. Albert, A. L. Barabasi \textit{Statistical mechanics of complex networks},
Reviews of Modern Physics \textbf{74}, 47-97 (2002).


\bibitem{a18} A. Barra,
\textit{The mean field Ising model trhought interpolating
techniques}, J. Stat. Phys. \textbf{132}, (2008).

\bibitem{a19} A. Barra,
\textit{Irreducible free energy expansion and overlap locking in
mean field spin glasses}, J. Stat. Phys. \textbf{123}, (2006).

\bibitem{a20} A. Barra, F. Guerra, \textit{About the ergodicity in Hopfield analogical neural
network}, J. Math. Phys. \textbf{50}, (2008).


\bibitem{a22} A. Barra, F. Camboni, P. Contucci, {\em Structural stability of dilution in mean field
models}, J. Stat. Mech., P03028, (2009).

\bibitem{a23} A. Barra, G. Genovese, {\em An analytical approach to mean field systems defined on
lattice}, J. Math. Phys. \textbf{51}, (2009).

\bibitem{a24} A.C.C. Coolen, R. Kuehn, P. Sollich, {\em Theory of Neural Information Processing
Systems}, Oxford Univ. Press, (2005).

\bibitem{a25} H. Jacobs, L. Bross, {\em Towards an understanding of somatic
hypermutation}, Current Opinion in Immunology \textbf{13},
208-218, (2001).

\bibitem{a26} U. Storb, {\em The molecular basis of somatic hypermutation of immunoglobulin
genes}, Curr. Op. in Imm. \textbf{8}, 206-214, (1996).

%
%

\bibitem{a30} I. Lundkvist, A. Coutinho, F. Varela, D. Holmberg,
 {\em Evidence for a functional idiotypic network among natural
 antibodies in normal mice}, P.N.A.S.  \textbf{86},
 $13:5074-5078$, (1989).

%

\bibitem{a33} M. M\'ezard, G. Parisi and M. A. Virasoro, {\em Spin glass theory
and beyond}, World Scientific, Singapore (1987).

\bibitem{a34} J.M. Seigneurin, B. Guilbert, M.J. Bourgeat, S. Avrameas, {\em Polyspecific natural
antibodies and autoantibodies secreted by human lymphocytes
immortalized with Epstein-Barr virus }, Blood, \textbf{71}, 3,
581-585, (1988).

\bibitem{a35} G. Parisi, {\em A simple model for the immune
network}, P.N.A.S. \textbf{87}, $(1990)$.

\bibitem{a36} P. Pereira, L. Forni, E.L. Larsson, M. Cooper,
C. Heusser, A. Coutinho, {\em Autonomous activation of B and T
cells in antigen free mice}, Eur. Journ.  Immun., \textbf{16},
(1986).


\bibitem{a38} J. Stewart, F. J. Varela, A. Coutinho, {\em The relationship between connectivity and tolerance as revealed
by computer simulation of the immune network: some lessons for an
understanding of autoimmunity}, J. of Autoimmunity, \textbf{2},
(1989).

\bibitem{a39} F.J. Varela, A. Countinho,
{\em Second generation immune networks}, Imm. Today \textbf{12},
 5, 159, (1991).


\bibitem{a40} Edelman, G.M., {\em The problem of molecular recognition by a selective
system}, In Ayala/Dobzhansky, Studies in the Philosophy of
Biology,  $45,56$, (1974).


\bibitem{a42} Ledemberg, {\em Genes and antibodies}, Nobel
Prize Lecture.

\bibitem{a43} I. Gallo, P. Contucci
{\em Bipartite mean field spin systems. Existence and solution }, Math. Phys. E. J. \textbf{14},  (2008).

\bibitem{a44} N.K. Jerne, {\em Toward a network theory of the
immune response}, Ann. Imm. \textbf{125}, C, (1974).


\bibitem{a48} F.M. Burnet, {\em A modification of Jerne's
theory of antibody production using the concept of clonal
selection}, Australian J. Science \textbf{20},, 67, (1957).


\bibitem{a50} A. Cooper-Willis, G.H. Hoffmann. {\em Symmetry of effector
function in the immune system network}, Mol. Imm. \textbf{20},
865, (1983).

\bibitem{a51} H. C. Tuckwell, {\em Introduction to theoretical
neurobiology}, Cambridge St. Math. Bio., Cambridge Press, (1988).

\bibitem{a52} L. DeSanctis, F. Guerra, {\em The dilute ferromagnet: high temperature and zero temperature}, J. Stat. Phys.
\textbf{132}, (2008).
%
%
\bibitem{a54} J.P.L. Hatchett, I. P\'erez Castillo, A.C.C. Coolen, N.S. Skantzos, {\em Dynamical replica analysis of disordered Ising spin systems on finitely connected random
graphs}, Phys. Rev. Lett. \textbf{95}, 117204, (2005).

\bibitem{a55} H. Lemke, H. Lange,{\em Generalization of single immunological
experiences by idiotypically mediated clonal connections}, Adv.
Immunol. \textbf{80}, 3078-3082, (1980).

\bibitem{a56} M. Molloy and B. Reed, {\em A critical point for random graphs with a given degree sequence}, Random Structures and Algorithms \textbf{6}, 161-180, (1995).

\bibitem{a57} A. Nachmias and Y. Peres, {\em Component sizes of the random graph outside the scaling window}, Alea \textbf{3}, 133-142, (2007).
%


\bibitem{a60} B.K. Chakrabarti, M. Acharyya, {\em Dynamic transitions and
hysteresis}, Rev. Mod. Phys. \textbf{71}, 3, (1999).

\bibitem{a61} K.F. Ludvig, B. Park, {\em Kinetics of true-field Ising models and the Langevin equation: A
comparison}, Phys. Rev. B, \textbf{46}, 9, (1992).

\bibitem{a62} J.O. Sethna, K. Dahmen, S. Kartha, J.A. Krumhansl, B.W. Roberts,
J.D. Shore, {\em Hysteresis and hierarchies: Dhyamics of
disorder-driven first-order phase transformations}, Phys. Rev.
Lett. \textbf{70}, 21, (1993).


\bibitem{a67} L.A. Segel, A.S. Perelson, {\em Shape space: an approach to the evaluation of cross-reactivity effects, stability
and controllability in the immune system}, Imm. Lett. \textbf{22},
91-100, (1989).

\bibitem{a68} J. Chun, {\em Selected comparison of immune and nervous system
development}, Adv. Imm. \textbf{77}, (2001).

\bibitem{a69} P.M. Colman, {\em Structure of antibody-antigen complexes: implication for immune
recognition}, Adv. Imm. \textbf{43}, (1988).

\bibitem{a70} P.A. Cazenave, {\em Idiotypic-anti-idiotypic regulation of antibody synthesis in
rabbits}, P.N.A.S. \textbf{74}, vol 11, 5122-5125, (1977).


%
\bibitem{a75} W.E. Paul, C. Bona, {\em Regulatory idiotopes and immune networks: a
hypothesis}, Imm. Today \textbf{3}, 9, (1982).



\bibitem{a79} R.J. De Boer, {\em Symmetric idiotypic networks: connectance and switching, stability and
suppression},  Theor. Immun. Vol.$2$, Studies in the sciences of
complexity, Add.-Wiley Publ. (1988).

%
\bibitem{a81} E. Agliari, A. Barra, F. Camboni, \textit{Criticality in diluted
ferromagnets}, J. Stat. Mech. P1003 (2008).

%
\bibitem{a89} R.S. Ellis,
\textit{Large deviations and statistical mechanics}, Springer, New
York (1985).

\bibitem{a90} F. Guerra, F. L. Toninelli, {\em The high temperature region of the Viana-Bray diluted spin glass
model}, J. Stat. Phys. \textbf{115}, (2004).

\bibitem{a91} M.Mezard, G.Parisi, R. Zecchina, \textit{Analytic and Algorithmic Solution of Random Satisfiability Problems},
 Science \textbf{297}, 812 (2002).


\bibitem{a93} M. Newman, D. Watts, A.-L. Barabasi
\textit{The Structure and Dynamics of Networks}, Princeton
University Press, (2006).

\bibitem{a94} A.C.C. Coolen, {\em The Mathematical Theory of Minority Games - Statistical Mechanics of Interacting
Agents}, Oxford University Press, (2005).
%

\bibitem{a95} Durlauf, S. N.: 1999, \textit{How can statistical mechanics contribute to social science?}.
Proceedings of the National Academy of Sciences of the U.S.A.
\textbf{96}, 10582-10584, (1999).

\bibitem{a96} Mc Fadden, D. (2001), \textit{Economic Choices}, The American Economic Review, 91: 351-378, (2001)

\bibitem{a97} M. M\'ezard, G. Parisi and M. A. Virasoro, {\em Spin glass theory
and beyond}, World Scientific, Singapore (1987).

\bibitem{a98} G. Semerjian, M. Weigt {\em Approximation schemes for the dynamics of
diluted spin models: the Ising ferromagnet on a Bethe lattice} J.
Phys. A \textbf{37}, $5525$ (2004).
%

\bibitem{a100} A. Barra, G. Genovese, {\em A certain
class of Curie-Weiss models}, Math. Phys. El. J. (2010).

\bibitem{a101} P. Delves, S. Martin, D Burton, I Roitt,
{\em Roitt's Essential Immunology}, Wiley Ed.r, (2006).

\bibitem{a102} P. Sollich, {\em Soft glassy rheology}, Phys. Rev. E

\bibitem{a103} N.R. Rose, I.R. Mackay, {\em The autoimmune
diseases}, Elsevier Press, (2006).

\bibitem{a104} F. Mandl, G. Shaw, {\em Quantum field theory}, Wiley
Ed.r, (1984).

\bibitem{a105} J. Monod, {\em Chance and Necessity}, A. Knopf Ed.r, New York,
(1971).

\bibitem{bagley} R. J. Bagley, J.D. Farmer, N.H. Packard, A.S. Perelson, I.M. Stadnyk, {\em Modeling adaptive biological systems},  Biosystems. \textbf{23}, 113-138, (1989).

\bibitem{farmer} J. D. Farmer, A. Lapedes, N.H. Packard, B. Wendross, {\em Evolution, Games and Learning}, eds. North-Holland, (1987).

\bibitem{a106} S.C. Bagley, H. White, B.A. Golomb, {\em Logistic regression in the
medical literature},  J. Clin. Epidem. \textbf{54}, 10, (2001).

\bibitem{a107} D. Frenkel, B. Smith, {\em Understanding molecular
simulation}, Elsevier Press, (2002).

\bibitem{a108} R.J. Be Boer, I.G. Kewrekidis, A.S. Perelson {em Immune network
behavior. From stationary to limit cycle oscillations}, Bull.
Math. Biol, \textbf{55}, 745-780 (1993).

\bibitem{a109} G,W, Hoffmann, T.A. Kion, R.B. Forsyth, K.G.
Soga,A. Cooper-Willis, {\em The $N$-dimensional network}, Theor.
Imm. \textbf{2}, A.S. Perelson Ed.r, Addison-Wiley Pubbli.,
(1988).

\bibitem{holmberg} D.Holmberg, G. Wennerstr\"{o}m, L. Andrade and A. Coutinho {\em The high idiotypic connectivity of `natural' newborn antibodies is not found in adult mitogen-reactive B cell repertoires}, Eur.\
J.\ Immunol.\, \textbf{16}, 82-87 (1986).

\bibitem{per_string} V. Detours, B. Sulzer and A. Perelson, {\em Size and Connectivity of the Idiotypic Network Are Independent of the Discreteness of the Affinity Distribution}, J.\ Theor.\
Biol. \textbf{183}, 409-416,
(1996).



\end{thebibliography}
\end{document}